\documentstyle[11pt,aaspp4]{article}

\epsfclipon

\lefthead{Lopez et al.}
\righthead{Ionization and Metallicity in Two Lines of Sight}

\begin{document}

\title{First Comparison of Ionization and Metallicity in Two Lines of
Sight Toward HE 1104--1805 AB at \boldmath
$z=1.66$\unboldmath\footnote{
Based on observations made at ESO, La Silla,
Chile. Based on observations made
at the Anglo Australian Telescope. Based on
observations with the NASA/ESA {\it Hubble Space Telescope}, obtained at the
STScI, which is operated by AURA, Inc., under NASA contract
NAS5--26\,555. Based on observations made at the W. M. Keck
Observatory, which is operated as a scientific partnership between the
California Institute of Technology and the University of California;
it is made possible by the generous support of the W. M. Keck Foundation.
}}

\author{Sebastian Lopez and Dieter Reimers}
\affil{Hamburger Sternwarte, Universit\"at Hamburg, Gojenbergsweg 112,
21029 Hamburg, Germany; slopez@hs.uni-hamburg.de}

\author{Michael Rauch\footnote{Hubble Fellow} and Wallace L. W. Sargent}
\affil{Astronomy Department, 105-25 California Institute of Technology,
1200 E. California Blvd., Pasadena, CA 91125, USA; mr@astro.caltech.edu}

\and

\author{Alain Smette\footnote{Now at Laboratory of Astronomy and Solar Physics,
NASA-Goddard Space Flight Center, Code 681, Greenbelt MD 20771, USA, and
National Optical Astronomy Observatories, P.O. Box 26732, 950 N. Cherry
Ave., Tucson, AZ 85726-6732; asmette@band3.gsfc.nasa.gov}}
\affil{Kapteyn Astronomical Institute, PO Box 800, NL-9700 AV
Groningen, The Netherlands.}

\begin{abstract}
Using new {\it Hubble Space Telescope} Faint Object Spectrograph, {\it
New Technology Telescope} EMMI, and Keck HIRES spectra of 
the gravitationally-lensed double QSO HE 1104--1805 AB
($z_{\rm em}=2.31$), and assuming UV photoionization by a metagalactic
radiation field,
we derive physical conditions (ionization levels, metal abundances and
cloud sizes along the lines of sight) in five C\,{\sc iv}+Mg\,{\sc ii}
absorption systems
clustered around $z=1.66$ along the two lines of sight. 
Three of these systems are associated with a damped Ly$\alpha$ (DLA) 
system with log\,$N$(H\,{\sc i}) = 20.85, which is 
observed in the ultraviolet spectra 
of the bright QSO image (A). The other two systems are associated 
with a Lyman-limit system with log\,$N$(H\,{\sc i}) =
17.57, seen in the fainter image (B). 
The C\,{\sc iv} and Mg\,{\sc ii} line profiles 
in A resemble those in the B spectra, 
and span $\Delta v\approx 360$ km~s$^{-1}$.
The angular separation $\theta=3.195$\arcsec{} between A and B corresponds
to a transverse proper separation of $S_{\bot}=8.3$~h$_{50}^{-1}$
kpc, for $q_0=0.5$ and a lens at $z = 1$. 

Assuming that the {\it relative}
metal abundances in these absorption systems are the same 
as observed in the DLA system, 
we find that the observed $N$(C\,{\sc iv})/$N$(Mg\,{\sc ii}) 
ratios imply ionization parameters
of $\log\Gamma=-2.95$ to $-2.35$. Consequently, these clouds 
should be small (0.5--1.6 kpc with a hydrogen density
$n_{\rm H}$ \rlap{\lower 3.5 pt \hbox{$\mathchar \sim$}}\raise 1pt
\hbox {$<$} $0.01$ cm$^{-3}$) and relatively 
highly ionized. 
The absorption systems to B are found to have a metallicity 0.63 times
lower than the metallicity of the gas giving rise to the DLA system, 
$Z_{\rm DLA}\simeq 1/10~Z_\odot$. 

We detect O\,{\sc vi} at $z=1.66253$ in both QSO spectra, but no
associated N\,{\sc v}. Our model calculations lead us to conclude that
the C\,{\sc iv} clouds should be  surrounded by large ($\sim 100$ kpc) 
highly ionized low-density clouds 
($n_{\rm H}\sim 10^{-4}$ cm$^{-3}$), in which 
O\,{\sc vi}, but only weak C\,{\sc iv} absorption occurs. 
In this state, $\log\Gamma\geq-1.2$ reproduces the observed ratio 
of $N$(O\,{\sc vi})/$N$(N\,{\sc v}) $>$ 60. 

These results are discussed in view of the disk/halo and 
hierarchical structure formation models.

\end{abstract}


\section{Introduction}
\label{sec_int}

The physical nature of the ionized
gas observed in high redshift QSO absorption systems has not been clearly 
determined so far. 
One possible mechanism capable of ionizing this gas is ultraviolet 
(UV) photoionization by the background field of distant active galactic
nuclei and of local sources  associated with the absorption systems. 
In this model, two gas phases with different ionization 
parameters $\Gamma$ (ratio of hydrogen ionizing photon density to total 
hydrogen density $n_{\rm H}$) are commonly invoked in order to
explain the presence of low and high ionization species, e.g., 
Mg\,{\sc ii}, C\,{\sc iv}, and O\,{\sc vi}
(Bergeron et al.~\cite{Bergeron}; Lu \& Savage \cite{Lu}). As $\Gamma$ is
defined by $n_{\rm H}$ for a given ionizing field, one can also estimate the
spatial extent of the ionized clouds along the line of sight (LOS) if the total
hydrogen column density $N$(H) is known. In general, the reliability of these
models strongly depends on a good knowledge of the metal abundances
involved. On the other hand, double LOSs toward background QSOs 
offer a unique possibility to resolve such
clouds geometrically and determine {\it transverse} sizes. 
For instance, Smette et al. (\cite{Smette1}) report lower limits of $D=0.7$ to
2.2~h$_{50}^{-1}$ kpc for the diameters of two metal systems at
$z\approx 2$ toward UM 673 AB, from the analysis of the line strengths
in both QSO spectra. A somewhat different result, 
although with large uncertainties, is achieved through statistical
simulations by Smette et al. (\cite{Smette}, hereafter Paper I), who infer 
$25<D<300$~h$_{50}^{-1}$ kpc for
C\,{\sc iv} clouds in the LOSs to HE 1104--1805 AB. 
At higher resolution, Rauch (\cite{Rauch1}) has
recently shown that density
gradients on sub-kpc scales in gas associated with metal systems are
not uncommon. These results suggest that C\,{\sc iv} absorbers are
composed of a large number of small cloudlets.

HE 1104--1805 AB ($z_{\rm em}=2.31$, $m_{\rm B}(A)=16.7$, 
$m_{\rm B}(B)=18.6$, angular separation $\theta=3.195$\arcsec{}) was discovered
in the course of the Hamburg/ESO Survey (Wisotzki et
al. \cite{Wisotzki1}), and has been already studied at medium
resolution by Smette et
al. (Paper I). There is wide evidence for this QSO to be gravitationally lensed
(Paper I; Wisotzki et al. \cite{Wisotzki2}; Courbin, Lidman, \& 
Magain~\cite{Courbin}), so
the proper separation between LOSs will depend on the (hitherto not
well established) redshift of the lensing agent. 
Two papers aimed at detecting the lensing galaxy
of HE 1104--1805 with somewhat different results have recently appeared.
Using IR direct imaging observations, 
Courbin et al. (\cite{Courbin}) estimate the lensing
galaxy to be at $z_{\rm lens}=1.66$, while Remy et al. (\cite{Remy}) 
find that their
{\it Hubble Space Telescope} ({\it HST}) and ground-based direct
imaging observations are consistent with
$z_{\rm lens}=1.32$. Neither of these results influences our 
photoionization models. Nevertheless, the small velocity differences
and the similar line profiles between absorption lines in A and B at $z=1.66$ 
(see \S~\ref{sec_geo}) probably exclude the damped Ly$\alpha$ (DLA)
system observed in A as the lensing agent, 
thus excluding the $z_{\rm lens}=1.66$ result.

The primary aim of this study is to compare ionization conditions and
metallicities in clouds at $z_{\rm abs}=1.66$ for the two LOSs toward
HE1104--1805. We use new {\it HST} Faint Object Spectrograph (FOS),
and ground-based  
{\it New Technology Telescope} ({\it NTT}) and Keck spectra. The
C\,{\sc iv} systems observed in 
the UV and optical spectra of A and B 
at redshifts close to $z_{\rm DLA}$ are 
very well suited to determine whether this gas is 
indeed photoionized: it can be assumed that they are associated
with the high-density gas giving rise to the DLA system observed in A 
at $z=1.66162$ and have therefore a common chemical history.
Consequently, the {\it relative} element abundances should be the same in
all these systems. Since the DLA gas is expected to be opaque to the 
ionizing radiation, it is possible 
to derive these element abundances without ionization corrections. 

To investigate the issue of ionization in the C\,{\sc iv} systems, 
we have used the photoionization 
code CLOUDY (Ferland \cite{Ferland}) and 
the ionizing radiation field proposed by
Haardt \& Madau (\cite{Haardt}). 
We have assumed that this radiation field also ionizes
the gas that gives rise to the strong O\,{\sc vi} absorption observed
along both LOSs. We have made a distinction
between two gas phases with different ionization states: a low
ionization phase, where both Mg\,{\sc ii} and C\,{\sc iv} absorption
occur, and a high ionization phase, where O\,{\sc vi} but weak C\,{\sc
iv} absorption occurs.
The photoionization models for each of these two gas phases are constrained by
the observed column density ratios $N$(C\,{\sc iv}) to $N$(Mg\,{\sc
ii}) and  $N$(O\,{\sc vi}) to $N$(N\,{\sc v}), respectively. 

Our paper is organized basically in two parts: estimation of the 
metal abundances in the
DLA gas, and photoionization models for the C\,{\sc iv} systems. 
The spectra are described in \S~\ref{sec_obs}. In
\S~\ref{sec_abs} we describe the line fitting method used, 
and discuss the important role played by the continuum fitting 
to the {\it HST} spectra. \S~\ref{sec_abu} is devoted to the metal abundances
determined for the DLA system observed in QSO component A. 
\S~\ref{sec_geo} presents possible physical
scenarios for the C\,{\sc iv} systems observed in A and B based on their
association with the DLA system and on the line profiles. The CLOUDY models
and the resulting physical parameters for the C\,{\sc iv} and O\,{\sc
vi} absorbers are described in \S~\ref{sec_ion}. 
Finally, we outline our conclusions in \S~\ref{sec_con}.

\section{Observations and Data Reduction}
\label{sec_obs}

An overview of the spectra used in this paper is displayed in
Table~\ref{tbl-5}. We now detail the observations.

\placetable{tbl-5}

\subsection{{\it HST} Spectra}

UV spectra of HE 1104--1805 A and B were taken in November 1995 with the
Faint Object Spectrograph onboard the {\it Hubble Space 
Telescope}. Target acquisition and spectroscopy were done
using Grating G270H with the red detector and the $3.\!{}''7\times3.\!{}''7$
aperture. This configuration
yields a spectral resolution of FWHM $=2$ \AA{} and a wavelength coverage from
2222 \AA{} to 3277 \AA{} (Schneider et al. \cite{Schneider}). Total
integration times of 1790 and 6690 seconds for QSO component A and B,
respectively, resulted in variance weighted spectra of  
maximum signal-to-noise ratios S/N $=20$ (A) and $17$ per $\sim 0.5$ \AA{}
pixel.

\subsection{Keck HIRES Spectra}

Optical spectra of HE 1104--1805 A and B were 
taken in January and February 1997 with the Keck High Resolution
Spectrograph HIRES (Vogt et al.~\cite{Vogt}) and an 0.86\arcsec{} slit
at FWHM = 6.6 km~s$^{-1}$. They 
range from 3620 to 6080 \AA. 
A full description of the observations and extraction method, as well
as the spectra themselves will be presented elsewhere. In short, 
an image rotator was used to keep the
slit off the second image, and as close as possible to the parallactic
angle. The continua were matched by using the brighter A image
continuum as a template, i.e., the continuum points of the B image 
spectrum were, in a sense, "stapled" to the A image continuum. This was
done using polynomial fits in such a way that differences
between the continua on scales larger than at average 300 km~s$^{-1}$ are
divided out, but regions smaller than that retain their differences
between the LOSs.  The typical absorption line or absorption complexes
between the spectra (which were omitted from the fit anyway) are not 
affected as can be seen from the strong differences in the metal
lines despite the very similar Ly$\alpha$ forest. We have also taken
special care to omit the metal absorption line complexes from 
the continuum points,
to be sure that we do not wipe out the differences.
One-sigma arrays were derived from the Poissonian
photon error, and rebinned onto the 0.04 \AA/pixel constant wavelength scale.

\subsection{{\it NTT} Spectra}

Optical spectra of HE 1104--1805 A and B were obtained in
February 1996 with the echelle spectrograph of EMMI 
on the ESO {\it New Technology Telescope} in the
Red Medium Dispersion mode under subarcsecond seeing conditions. 
The wavelength coverage is 3910 to 8290 \AA. 
Both images were simultaneously centered in a 
$1''$ slit. Grating {\#}9, grism 3 as cross-disperser, 
a F/5.2 camera and a TEK $2048^{2}$ CCD were used. 
This CCD provides $24 \mu$m pixels, corresponding 
to $0.\!{}''27$ in the sky. The total
integration time was 16 hours. 
The echelle orders 
were optimally extracted with a
version of the extraction algorithm used in Paper I,
modified to extract cross-dispersed spectra. The algorithm attempts to
reduce the statistical noise in the extracted spectra to a minimum,
and allows the correct separation of the two seeing profiles. It
basically consists of the following steps: (i) 
a variance (Poisson statistics, read-out noise and
cosmics) is assigned to each pixel. For each flat-fielded
two-dimensional spectrum, two Gaussians of common
width are simultaneously fitted to the profiles at each wavelength
channel along the previously defined echelle orders using the
Levenberg-Marquardt method (Press et al. \cite{Press}); (ii) 
the variation of the width and the position of the brighter component
with respect to the orders 
in the dispersion direction is then fitted with low order
polynomials; (iii) step (i) is repeated, this time with fixed width and
position---given by the polynomial fits---thus allowing only the
amplitudes to vary. The final, variance-weighted coadded spectra
have FWHM $=0.8$ to $1.0$ \AA{} and 
maximum S/N $\simeq 77$ (A) and $\simeq 27$
(B). The continuum level was estimated separately 
in each order---skipping corrections for the blaze function---and  
for each QSO image. This was done by fitting low-order polynomials and
cubic splines to featureless spectral regions.

\subsection{{\it AAT} Spectra}

Additional FWHM $=1.2$ \AA{} resolution spectra of both QSO images 
taken with the 3.9m {\it Anglo-Australian Telescope} and covering the 
wavelength range 3170 to 7570 \AA{} were also
used. They have already been presented in Paper I.

\subsection{Wavelength calibration of the {\it HST} spectra}

Special care has been taken to re-define the absolute zero point of
the wavelength scale in the {\it HST} spectra.
An off-center position of the targets in the aperture of the 
FOS may cause differences in the wavelength scale between the A and B 
spectra. By measuring wavelength positions of the galactic
Mg\,{\sc ii} $\lambda\lambda2796,2803$  lines, and assuming that this
absorption takes place in the same cloud along both LOSs 
we found an offset of $\Delta\lambda(A-B)=-0.42$ \AA{}. This correction
was applied to the B spectrum. 
Additionally, an overlapping 100 \AA{} wide
spectral region around $\lambda$ = 3220 \AA{} in the {\it HST} and {\it AAT}
spectra allowed for a correction of
the FOS wavelength scale to vacuum-heliocentric values. 
This was done by comparing the central wavelength differences of 
absorption lines in the A spectrum thought not to be blended. 
A relatively small correction of $\Delta\lambda({\it AAT}-{\it HST})=+0.04$
\AA{} was then applied to both {\it HST} spectra.

\section{Absorption Line Analysis}
\label{sec_abs}

\subsection{Continuum Definition in the {\it HST} Spectra}

Owing to the low FOS resolution, line blending introduces a serious 
problem when performing continuum fitting. This is particularly marked
in the Ly$\alpha$ forest, because of the lack of 
spectral regions free of absorption lines. For this reason, we decided
to first determine the
{\it absorbed} continuum---clearly dominated by two optically
thin Lyman-limit systems (LLS) 
at $z=2.20$ and $2.30$ and the DLA system (A) and a LLS at
$z=1.66$ (B)---before describing the intrinsic QSO continuum.
Both spectra were corrected for galactic extinction (Seaton
\cite{Seaton}) with E(${\rm B}-{\rm V}$)=0.09 (Reimers et al. \cite{Reimers}).
The Lyman edges, arising from the superposition and blending of 
corresponding high-order H\,{\sc i} Lyman series lines, were
modeled with Voigt profiles convolved with a FWHM = 2 \AA{} Gaussian
describing the instrumental profile. The optical depth $\tau$ at
each Lyman break, given by the ratio of the extrapolated continuum to
the Lyman one, determined $N$(H\,{\sc i}),
whereas $b$ was mainly constrained by the shape of the edge. These
parameters are listed in Table~\ref{tbl-1}. 
The total optical depth at each LLS is given by
the Lyman continuum and the lines of the Lyman series. Note 
that the flux for $\lambda<2200$ \AA{} in the B
spectrum is not completely absorbed (see Fig.~\ref{fig1}).

\placefigure{fig1}
\placetable{tbl-1}

The intrinsic QSO continuum of HE 1104--1805 A and B in the {\it HST} spectra 
for $\lambda_{\rm{obs}}\leq 3277$ \AA{} 
was found to be very well represented by two power laws 
($f\propto\nu^{\alpha}$) with a
break at $\lambda$ = 2917 \AA. To determine the power law parameters, 
a maximum likelihood fit was performed, 
matching simultaneously the intrinsic and the
H\,{\sc i} absorption continuum  with the observed flux at 
selected absorption-free spectral 
regions including some regions dominated by significant 
H\,{\sc i} absorption lines. This led to the best-fit spectral 
indices $\alpha=-0.8$, $-1.5$ for A, 
and $-0.8$, $-0.9$ for B (see Fig.~\ref{fig1}).
We believe this continuum estimation is a good one, since an extrapolation to
longer wavelengths fits the scaled {\it AAT} spectra to well within 1$\sigma$
in regions of low Ly$\alpha$ line density. Moreover, expected
emission lines by Ly$\alpha$, N\,{\sc v}, and O\,{\sc vi} stand out
well against the continuum as can be seen in Fig.~\ref{fig1}. Division
of the flux by this continuum resulted in the normalized {\it HST}
spectra shown in Fig.~\ref{fig8}.

\placefigure{fig8}

The different continuum slopes of A and B, already pointed out in the
discovery paper, might be a consequence of QSO component A being
microlensed (Wisotzki et al. \cite{Wisotzki1}). Furthermore, we find the
spectrum of A bluewards of Ly$\alpha$ emission 
to be softer than that of B,
in agreement with the variability in the spectral slopes reported by
Wisotzki et al. (\cite{Wisotzki2}) from the analysis of low resolution
observations in the optical range made one and two years before
ours. However, 
an alternative explanation for the different slopes might be
differential reddening by dust grains in the DLA system observed in A, a
possibility recently considered for the gravitationally lensed QSO
0957+561 (Zuo et al.~\cite{Zuo}).

\subsection{Line Profile Fitting}
\label{sec_col}

In this section we describe the line fitting procedures used to 
obtain column densities of lines associated with the $z=1.66$
absorption systems in A and B. In general, this is a nontrivial task
because of the limited resolution of our spectra other than Keck
HIRES, so that assumptions concerning line widths must be made. The 
lines were modelled with Voigt profiles. We 
distinguish between maximum likelihood fits, performed to lines in
the Keck and {\it NTT} spectra, and ``interactive'' fits, performed to
lines in the {\it
AAT} and {\it HST} spectra (aside from H\,{\sc i} lines).

\subsubsection{Keck Spectra}
\label{sec_lin}

To determine column densities $N$ and Doppler parameters $b$ of lines
in the Keck HIRES spectra, 
we $\chi^2$-fitted Voigt profiles convolved with the instrumental
profile to lines in 
the Keck spectra using the MIDAS program FITLYMAN (Fontana \& Ballester 
\cite{Fontana}). These lines lie longward of Ly$\alpha$
emission. Line parameters, i.e., rest-frame vacuum wavelengths, 
damping constants and oscillator strengths, were taken 
from Morton (\cite{Morton}), and from Verner et al. (\cite{Verner})
for lines with revised $f$-values.

Most of the column density errors $\sigma_{{\rm log}N}$ range from 0.01 to 0.10
dex; however, the smoothing introduced by rebinning the
1$\sigma$-arrays sometimes 
underestimates the true flux uncertainties, making $\sigma_{{\rm log}N}$
systematically too low. Thus, we made the following  correction to the
fit errors: for each line, we measured the amount of smoothing by
calculating the ratio of the flux standard deviation  from 
41-pixel wide featureless regions in
the data to the sigma from the error array. In the (few) cases where this
ratio was larger than one, $\sigma_{{\rm log}N}$ was corrected by this
factor. Clearly, this applies only to unsaturated,
resolved lines; a similar treatment to non-resolved complexes is less
obvious because the $\sigma_{{\rm log}N}$'s result from a more complicated
Hessian matrix, and are no longer independent. We did not allow for 
this effect. 

Special attention must be given to the fits of lines associated with
the DLA system observed in the spectrum of QSO component A because they will
determine the metal abundances. To look for possible hidden
saturation, we used the apparent optical depth method to compare apparent
column densities $N_{\rm app}(v)$ of different transitions 
of a given ion in velocity space (e.g. eq. [1] in Lu et
al.~\cite{Lu1}; Savage \&{} Sembach~\cite{Savage1}, for a description of
the method). Although some lines are probably saturated
(e.g. Al\,{\sc ii} $\lambda1670$), in most of the cases where more
than one transition is available we do not
find significantly saturated structures, and the agreement 
between the fit and integration results is remarkable 
(cf. Table~\ref{tbl-3}).

\subsubsection{{\it NTT} Spectra}

In the {\it NTT} spectrum, the Mg\,{\sc ii} profiles at $z_{\rm DLA}$ 
are slightly asymmetric, suggesting that this system will also split
into more components at higher resolution. 
Consequently, we fitted four-component Voigt profiles to these lines, 
with $b$-parameters fixed at values found for the Si\,{\sc ii}
lines in the Keck spectra.

\subsubsection{{\it HST} and {\it AAT} Spectra}

At even lower resolution, one cannot expect to recognize the 
line profiles properly; hence,
to derive column densities of lines 
in the {\it HST} and {\it AAT} spectra, 
turbulence dominated line broadening was considered. 
Voigt profiles were interactively created and superimposed to the
spectra using XVOIGT (Mar \& Bailey~\cite{Mar}), while attempting to
minimize the residuals.
The redshift and Doppler widths used to create such line profiles were
those found 
in a {\it second} fit with FITLYMAN of lines present both in the {\it
AAT} and Keck spectra. Lines with 
asymmetric profiles were considered to be {\it single}, and the total
column densities were fixed to the known Keck-values.
In this fashion, the low and high-ionization species 
were distinguished using ``low-resolution'' Doppler
parameters determined by the fits to Fe\,{\sc ii} $\lambda1608$ 
($b=20$ km~s$^{-1}$) 
and C\,{\sc iv} $\lambda1548$ ($b=44$ km~s$^{-1}$) lines in A, respectively. 
For B, the lines used were Al\,{\sc ii} $\lambda1670$ ($b=20$
km~s$^{-1}$) and C\,{\sc iv} $\lambda1548$ ($b=37$ km~s$^{-1}$).

To estimate the uncertainties of our column densities we smoothed and
rebinned lines observed in the {\it NTT} spectra to {\it HST} FOS resolution, and
re-computed column densities with the procedure described above. The
new column densities showed deviations of the order 0.1 to 0.2 dex
from the original, better determined values.
Another source of error
is our limited ability to de-blend metal lines from 
Ly$\alpha$ forest lines. On the other hand, most column densities of
transitions in the UV that contribute to the metal abundances 
are based on one line in the {\it HST} spectra and another
in the {\it AAT} ones, e.g., C\,{\sc ii} $\lambda\lambda$1036,1334; O\,{\sc i} 
$\lambda\lambda$988,1302 (see Fig.~\ref{fig8}).
If these two effects compensate, we think that taking
$\sigma_{\rm{log}N}=0.2$ dex for these ions is  
appropriate. 

\subsubsection{Detection Limits}

We defined 3 $\sigma$ detection limits for
metal lines in the {\it HST} and {\it AAT} spectra 
according to the formula (Caulet~\cite{Caulet})
\begin{equation}
\sigma_{W}=\frac{{\rm FWHM}}{<{\rm S/N}>},
\end{equation}
where FWHM is the width of the spectral point spread function and $<$S/N$>$
is the mean local signal to noise expected at the position of the line
(measured as the inverse standard deviation of the normalized flux in 
small featureless stretches adjacent to the line).
They range between $W_{\rm obs}=0.13$ and $1.00$ \AA{} in the B spectra.

\subsubsection{{\rm H\,{\sc i}} Column Densities at $z=1.66$}
\label{sec_hi}

To derive more accurate column densities for H\,{\sc i}, we used the
normalized {\it HST} spectra.
Because they completely cover the rest-frame spectral range
down to 912 \AA{} for the $z=1.66$ systems, it is possible to measure  
$N$(H\,{\sc i}) in both spectra 
more accurately than in previous studies on damped 
systems, by using higher Lyman series transitions. 
We simultaneously fitted two-component Voigt profiles to 
11 resolved H\,{\sc i} lines in the normalized spectra of 
A and B using FITLYMAN. The fit solutions to 
lines in B were constrained by $\tau=2.34$ at the Lyman edge. 
We estimate the neutral hydrogen column density of the DLA gas (spectrum A)
to be log\,$N$(H\,{\sc i}) = 20.85$\pm$0.01; for the LLS (B) we obtained
log\,$N$(H\,{\sc i}) = 17.57$\pm$0.10. Notice that
{\it these values are independent of the ones estimated for
placing the continuum and shown in Table~\ref{tbl-1}}.

\section{The Damped Ly$\alpha$ System Toward HE 1104--1805 A at \boldmath 
$z_{\rm DLA}=1.66162$}
\label{sec_abu}

Table~\ref{tbl-2} displays the fit results for lines associated with
the DLA system toward HE 1104--1805 A at $z_{\rm DLA}=1.66162$. 
A wide variety of singly and doubly ionized species is observed in
this DLA system, but C\,{\sc iv} and Si\,{\sc iv} are also present. We now 
describe the Keck HIRES line profiles, referring to 
the left-hand panels of Figures~\ref{fig9},~\ref{fig10} and~\ref{fig11},
throughout this section.

\placetable{tbl-2}

\subsection{Low Ion Profiles}

The left-hand panel of Fig.~\ref{fig9} shows the line profiles of strongest 
low-ionization species in velocity space, relative to $v=0$ 
at $z=1.66164$, the redshift of Mg\,{\sc ii} in the DLA system. These lines lie
redward of the Ly$\alpha$ forest. In each line complex, we have fitted
four Voigt profiles with {\it independent} $z$, $b$ and $N$ 
values. The Ni\,{\sc ii} and Fe\,{\sc ii} results are based upon 
simultaneous fits to 3 transitions; the 
Si\,{\sc ii} and Al\,{\sc iii} fits on 2 transitions; the Al\,{\sc ii}
fit on 1 transition; and the Mg\,{\sc ii} fit on 2 transitions 
(cf. Table~\ref{tbl-2}). 
We fitted 4 components to the Mg\,{\sc ii} lines in the 
{\it NTT} spectra, with $z$ and $b$ tied to the values found for Si\,{\sc ii}.
Only the three bluemost components show associated Zn\,{\sc ii} and 
Cr\,{\sc ii} (see~\ref{sec_zncr}).

\placefigure{fig9}

From the high-resolution plots, we see that the low ions track each
other quite closely, suggesting they occur in the same gas clouds. The whole
profile is characterized by one cloud at $v\approx-30$ km~s$^{-1}$
with the strongest absorption, one cloud at $v\approx+40$
km~s$^{-1}$ with the smallest column densities, and two clouds with 
intermediate column density clouds lying in between. This ``edge-leading  
asymmetry'' seems to be a common feature of low-ionization absorption
lines associated with damped Ly$\alpha$ systems. It has been variously
interpreted as a consequence of absorption by rotating gaseous disks
(e.g., Wolfe et al.~\cite{Wolfe1}) or as the signature of merging
protogalactic clumps in hierarchical structure formation (Rauch 
et al.~\cite{Rauch}). The different 
column density ratios at each velocity indicate clouds with different
physical conditions (gas density, metallicity, or even ionization) within
70 km~s$^{-1}$. 

\subsubsection{Line Widths}
\label{sec_wid}

Remarkably, and despite the {\it independent} fits performed to each ion
profile, we obtain fit solutions that uniquely characterize each cloud in
redshift and broadening parameter (cf. Table~\ref{tbl-2} and Fig.~\ref{fig9}). 
For instance, for the clouds at $v\sim-30$,
$-10$, $10$ and $40$ km~s$^{-1}$ we find respective mean widths and
standard deviations of
$\langle b\rangle =3.5\pm0.7$, $12.6\pm2.7$, $9.1\pm2.2$ 
and $7.6\pm1.8$ km~s$^{-1}$
(fourth Ni\,{\sc ii} component excluded due to large
$b$-uncertainties; fit to Zn\,{\sc ii}, Cr\,{\sc ii} and Ti\,{\sc ii} 
lines not included). Thus, we
find a similar line-broadening mechanism in each of these
absorption systems and, based on the small Doppler-parameter 
dispersions, cautiously favor turbulent gas motions as the dominant
line-broadening mechanism.

\subsection{High Ion Profiles}

The left-hand panel of Fig.~\ref{fig11} shows 
the C\,{\sc iv} $\lambda 1548,1550$,  Si\,{\sc iv} $\lambda 1393,1402$ and, 
for comparison purposes, Al\,{\sc ii} $\lambda 1670$ velocity profiles  
associated with the DLA system. For the C\,{\sc iv} absorption lines between 
[-80,+80] km~s$^{-1}$, only four-component Voigt profile fits succeeded. 
Unfortunately, the poor S/N at
the position of the Si\,{\sc iv} lines and contamination by 
Ly$\alpha$ forest lines 
do not allow a clear comparison with the C\,{\sc iv} profiles. We
decided not to fit these Si\,{\sc iv} lines. Instead, we give upper limits for
column densities based on the apparent optical depth method. 
The high ion profiles do not exactly track the
low ion profiles: the data rather suggest at least part of the C\,{\sc iv}
absorption (velocity component 1) occurs in clouds without low 
ionization species.
On the other hand, C\,{\sc iv} velocity components 2 to 4 show
a certain resemblance to the
edge-leading asymmetry of the low ion profiles; however, our fit
solution yields relatively large $b$-values, suggesting that C\,{\sc iv}, 
regardless of which line broadening mechanism
dominates---thermal or turbulent gas motion---occurs in hotter gas
regions than the low ion clouds.

\placefigure{fig11}

\subsection{Abundances}

Metal abundances in the DLA system, normalized to solar values and defined by 
\begin{equation}
{\rm [M/H]}\equiv\log (N({\rm M})/N({\rm H}))-\log(N({\rm M})/
N({\rm H}))_{\odot} 
\end{equation}
were computed using the column densities integrated between [-50,60]
km~s$^{-1}$. They are listed in
Table~\ref{tbl-3}. 
Solar abundances were taken from Verner et al. (\cite{Verner}).
Singly ionized species 
provide the bulk of the total element column densities, except for O\,{\sc
i}, assumed to be the dominant ionization stage of oxygen. 

\placetable{tbl-3}

\subsubsection{Zn, Cr and Ti Abundances and Dust}
\label{sec_zncr}

Fig.~\ref{fig10} shows the Keck velocity profiles of the most
outstanding Zn\,{\sc ii} and Cr\,{\sc ii} transitions. The fit
solutions lead to three-component profiles for the Cr\,{\sc ii}
complex at $v=-29.1$, $-7.3$ and $7.4$ km~s$^{-1}$, and two-component
profiles for Zn\,{\sc
ii} at $v=-30.0$ and $-9.0$ km~s$^{-1}$, relative to $z=1.66164$.
The column density ratios relative to solar vary from
$N$(Zn\,{\sc ii})/$N$(Cr\,{\sc ii}) $=3.47$ (bluemost component) to
$1.73$. If the Zn and Cr abundances as well as the ionization level were
the same in these clouds (which is very probable), 
then this variation  would indicate that the
dust-to-gas ratio in this DLA gas is inhomogeneous within 20 km~s$^{-1}$,
with a higher dust content in the higher density Zn\,{\sc ii}
component. 

\placefigure{fig10}

The variation of the abundance ratios of refractory elements among
different clouds associated with the DLA gas provide insights into the 
presence of dust in the disk of damped Ly$\alpha$ galaxies 
(Lu et al.~\cite{Lu1}), because dust grains can be locally destroyed
by passage of supernova shocks (Sembach \&{} Savage~\cite{Sembach}, 
and references therein). We can extend our analysis of elemental
abundance ratios to iron and nickel, also expected to be
depleted into dust. The column density ratios 
Zn\,{\sc ii} to Fe\,{\sc ii} and Zn\,{\sc ii} to Ni\,{\sc ii}
relative to solar are respectively Zn/Fe = 5.5 and Zn/Ni = 7.9 for 
velocity component 1, and Zn/Fe = 2.5 and Zn/Ni = 2.8 for velocity 
component 2, thus in concordance with
what one observes for Cr in the corresponding clouds (we estimate the
corresponding uncertainties to be no larger than 0.4). Although these
variations are small, $0.3$ -- $0.4$ dex, it seems that
in the $v\sim-30$ km~s$^{-1}$ cloud the effect of dust depletion is
more important than in the cloud at $v\sim-10$ km~s$^{-1}$. On the
other hand, as pointed out in section~\ref{sec_wid}, the bluemost
component is characterized by narrower lines than component 2 in all
the ions considered (including Zn\,{\sc ii} and Cr\,{\sc ii}). 
It would be of great interest to discern whether such line width
differences have a thermal origin (contrary to 
what was stated in~\ref{sec_wid}, however), 
because it would give evidence for dust
depletion being more effective in cooler gas.

Lu et al. (\cite{Lu1}) have presented arguments for a pure 
nucleosynthetic origin of the elemental abundance pattern 
observed in Ly$\alpha$ damped systems at low metallicity. 
These arguments are: the low N/O ratio,
which we also find in this DLA system (see next section); 
the $\alpha$-element overabundance relative to Fe-peak elements,
also observed for the abundance ratios of Si to Fe, Cr, Mn and Ni 
in our Keck HIRES data; and the underabundance of Al relative to Si and Mn
relative to Fe (the ``odd-even effect''; see Lu et al. \cite{Lu1} for
details), which we do not find in this DLA system. Instead, we derive
[Al/Si]$=+0.25\pm0.15$, and [Mn/Fe]$=-0.02\pm0.14$ 
(cf. Table~\ref{tbl-3}), although we recall that the Al abundance is 
only based on the Al\,{\sc ii} $\lambda 1670$ line. Given the high 
Mn/Fe ratio and the argument given in the last paragraph, 
we suggest that stellar 
nucleosynthesis {\it alone} is not likely to produce the relative abundance
pattern observed in this DLA gas. 

Based on the total zinc abundance [Zn/H] =
$-1.02\pm0.01$ relative to the solar value, we derive a
metallicity $Z_{\rm{DLA}}\simeq 1/10~Z_{\odot}$ for this system. This 
zinc abundance is somewhat lower than the value [Zn/H] $=-0.8$ 
reported by Pettini et al. (\cite{Pettini2}) [same data as in Paper
I], who also used log\,(Zn/H)$_{\odot}$ $=-7.35$, and more accurate since it is
based on a simultaneous fit to two Zn\,{\sc ii}
lines.\footnote{However, our $40$ km~s$^{-1}$ resolution {\it NTT} spectra
yield Zn and Cr abundances completely consistent with the Keck
results, thus validating abundance studies of Zn at 
medium resolution.} Additionally,
we deduce from [Cr/H] $=-1.46\pm0.02$ a dust-to-gas ratio of $\sim
0.11$, using the definition given by Vladilo (\cite{Vladilo}),
eq. [19], and considering that most of chromium should be incorporated 
into dust grains in
the ISM of galaxies showing damped H\,{\sc i} absorption (e.g. Pettini
et al. \cite{Pettini3}). Furthermore, titanium, another refractory element,
is found to have [Ti/H] $=-1.50\pm0.07$, based upon the unsaturated 
Ti\,{\sc ii} $\lambda\lambda1910.6,1910.9$ lines. 
This is in full agreement with the
incorporation of Ti and Cr into a dust phase in the neutral ISM of
this damped Ly$\alpha$ galaxy.
Thus, based on these abundances, we
conclude that (1) this DLA system does not differ too much from other ones
at higher redshifts, given the scatter observed in [Zn/H] (cf. Fig. 3
in Pettini et al.~\cite{Pettini2}); (2) there is evidence for
the presence of dust.

\subsubsection{The Abundance Ratio O/N}

The crucial abundance ratios in this work are [O/N] and [C/Mg]. 

Based on the O\,{\sc i} $\lambda\lambda$988,1302 lines, we find [O/H]
$=-0.98\pm0.20$, in very good agreement with the abundance ratio of
Zn. This supports O\,{\sc i} as a very good tracer of H\,{\sc i}---
as can be expected from its ionization potential, 13.62 eV---
and suggests that O is not depleted into dust in the ISM of this
damped Ly$\alpha$ galaxy, in agreement with observations in the local ISM
(Cardelli et al.~\cite{Cardelli}). Moreover, since both O and Si have been
observed to have solar abundance ratios in Galactic halo stars and in
metal-poor dwarf galaxies, one would expect [Si/O] $\simeq 0$ in DLA 
systems (Lu et al.~\cite{Lu1} and references therein). We do obtain the same
abundance ratios for O and Si within the errors, and [Si/H] is based on
reliable column-density measurements of two Si\,{\sc ii} lines
(observed in the Keck HIRES spectrum), making 
our oxygen abundance estimation yet more confident. Concerning the abundance
ratio of nitrogen, the bulk of [N/H] is provided by the
column density of N\,{\sc ii} $\lambda$1083. This line is very
probably {\it not} contaminated by a Ly$\alpha$ forest interloper, given
the absence of absorption at the same wavelength in the spectrum of 
B (see Fig.~\ref{fig8}). 
In addition, nitrogen---like oxygen---is also expected not to be 
depleted in the ISM (Cardelli et al.~\cite{Cardelli}). In consequence,
we are confident of an abundance ratio of [O/N]$=0.88$ in this DLA gas,
that is, O/N is 8 times greater than the solar ratio. 
This is in qualitative agreement with observations of damped Ly$\alpha$
galaxies at higher redshifts (Pettini et al.~\cite{Pettini1}; Lu et
al.~\cite{Lu1}), and with galactic chemical evolution models, 
because these two elements have different
nucleosynthetic origins, oxygen being produced in much shorter
timescales than nitrogen.

\subsubsection{The Abundance Ratio C/Mg}

The carbon abundance is based on the fits to the C\,{\sc ii} 
$\lambda\lambda1036,1334$ lines. 
C/Mg is found to have the solar value within the errors, 
but $N$(Mg) might be underestimated through saturation of 
the Mg\,{\sc ii} lines in A. 

\section{Geometry of the Absorbers at $z=1.66$}
\label{sec_geo}

Fig.~\ref{fig2} shows the velocity profiles of the C\,{\sc iv} 
$\lambda$1548,1550 and 
Mg\,{\sc ii} $\lambda$2796,2803 doublets (at $7$ and $40$ km~s$^{-1}$
resolution, respectively) toward HE1104--1805 A (left)
and B. In both panels, $v=0$ km~s$^{-1}$ corresponds to $z=1.66164$, 
which is the redshift of Mg\,{\sc ii} in the DLA system. We have arbitrarily
numbered the C\,{\sc iv} complexes at $z=1.66143$ (1), $z=1.66280$ (2)
and $z=1.66465$ (3) in the A spectra, and at $z=1.66184$ (4),
$z=1.66284$ (5) and $z=1.66493$ (6) in the B spectra, so they will be
refered to as systems 1 to 6 throughout the following sections. 
Also shown in Fig.~\ref{fig2}, covering a larger range of velocities,
are the profiles 
of the Ly$\alpha$, Ly$\beta$ and Ly$\gamma$ H\,{\sc i} lines. 
Note that H\,{\sc i} presents (at least) two components
both in A and B. 

\placefigure{fig2}

C\,{\sc iv} systems 1 to 4 show associated Mg\,{\sc ii}, although shifted
in velocity, as is more evident in systems 1 and 4 (the line
shapes suggest that either of these systems will probably split into more
components at even higher resolution). However, most
of the Mg\,{\sc ii} seen in A at $v=0$ is due to the DLA system.
In B, system 4 is identified with the LLS. The presence of Mg\,{\sc ii} 
in absorption systems 2 and 6 is less evident at this S/N, so only upper 
limits can be derived. C\,{\sc iv} system 5 will not be considered
here. 

Do C\,{\sc iv} and Mg\,{\sc ii} occur in the same clouds? Although 
the present data show a correspondence in velocity, there might not be
a physical association between both ions. However, we will show in the
next section that
photoionization simulations of these clouds do indeed predict
the presence of both ions in a common gas phase, in agreement with our
observations. Furthermore, we know from studies at high 
resolution that line profiles of low and high ions do track one
another in Lyman-limit systems (Prochaska \&{}
Wolfe~\cite{Prochaska1}). As discussed in \S~\ref{sec_high} it is also
possible that part of the C\,{\sc iv} 
arises in the same highly ionized gas that gives rise to O\,{\sc vi}. 
In any case, the velocity profiles of system 4
(LLS) show that, 
if a physical association of both ions is correct, at least part of
the C\,{\sc iv} arises in clouds where no Mg\,{\sc ii} is present
(see also the right-hand panel of Fig.~\ref{fig11}).

From inspection of Fig.~\ref{fig2}, it seems likely that LOSs A and B 
cross common absorbers. This is suggested by the similar line profile
pattern in both spectra. 
If a common absorption complex gives rise to systems 1 (A) and 4 (B), 
then the different equivalent widths of
the Mg\,{\sc ii} lines in A and B, in contrast to the more similar C\,{\sc iv}
equivalent widths, suggest gas inhomogeneities on spatial scales 
smaller than the linear separation between LOSs $S_{\bot}=8.3$~h$_{50}^{-1}$
kpc, for $q_0=0.5$ and a lens at $z=1$, thus suggesting that C\,{\sc iv}
arises in a more extended region than Mg\,{\sc ii}. 
However, we find that the velocity difference
between the C\,{\sc iv} clouds in A and B is 
$\langle\Delta v\rangle=-11\pm 2$ 
km~s$^{-1}$ while $\langle\Delta v\rangle=-1\pm 1$ km~s$^{-1}$ 
for the corresponding Mg\,{\sc ii} clouds. 
If these velocity differences are a consequence of
peculiar cloud motions, then a physical association of C\,{\sc iv} and
Mg\,{\sc ii} is still compatible with the data. 
Since it is not the aim of this study to determine the origin of both
this velocity difference and the $\sim$360 km~s$^{-1}$ velocity span
of the C\,{\sc iv} absorbers along both LOSs, we have analyzed each
system {\it separately}. 

There are two alternative interpretations for 
the H\,{\sc i} column densities in A and B: (1) the DLA system arises
in a disk-type galaxy (Wolfe \cite{Wolfe}) 
with the LOS to A passing through the gas in 
the halo and the disk and LOS B passing through the halo gas 
only, or (2) in models of  hierarchical structure formation the DLA
system arises in the central region of a protogalactic clump and LOS B
crosses the 
surrounding less dense gas at an impact parameter of a few kpc (Rauch
et al.~\cite{Rauch}). Our data do not allow us to discriminate between these
two models. Consequently, in the following we will simply consider
C\,{\sc iv} absorption systems 2 to 6 to arise in clouds
in the extended halo of the cloud giving rise to the DLA system
(system 1), giving explicit references to one of these models.

\section{Ionization State and Chemical Composition}
\label{sec_ion}

We have used the photoionization code CLOUDY (version 84.12a;
Ferland~\cite{Ferland}) to investigate the ionization state and
metallicity in the LLS observed in the B spectra of HE 1104--1805
(system number 4 in Fig.~\ref{fig2}) and in three further C\,{\sc iv}-Mg\,{\sc
ii} systems observed in A (systems 2, 3) and B (system 6). 
The clouds giving rise to these systems are represented by parallel
slabs illuminated on one side by the radiation field $J_\nu$ proposed by
Haardt \& Madau (\cite{Haardt}), consisting of the background flux
contributed by QSOs and AGNs, which is attenuated by H and He
absorption in Lyman-limit systems and Ly$\alpha$ forest clouds. This
radiation field also has a diffuse component due to 
recombination continuum radiation. 
Our model assumes $J(912)=0.37\times10^{-21}$ 
erg~s$^{-1}$~cm$^{-2}$~Hz$^{-1}$~sr$^{-1}$ at $z=1.66$. Although this
geometry does not perfectly describe a cloud being illuminated by an
isotropic incident radiation field, it should not introduce errors
larger than a factor of $\sim 2$ (Bergeron and
Stasinska~\cite{Bergeron1}). In particular, cloud sizes along the LOSs
resulting from this model need to be considered upper limits, because
considering slabs illuminated only on one side underestimates the true
ionizing radiation field.

As we argue below, the wide variety of low- and high-ionization 
stages present in these systems makes it necessary to model the gas
clouds with two zones of different ionization levels: (1) a
``low-ionization phase'', where absorption by singly, doubly, but
also triply ionized atoms occurs\footnote{The term ``low'' has 
specifically the purpose to distinguish both gas phases.}; 
 (2) a ``high-ionization phase'', where
O\,{\sc vi} absorption occurs. However, we will also discuss the possible
existence of a third, ``intermediate-ionization phase''.

The CLOUDY simulations predict column densities
of the expected ionization stages, suitable to be compared with the
observations. As a general strategy, we attempt to 
reproduce the observed ionic column density ratios 
$N$(C\,{\sc iv}) to $N$(Mg\,{\sc ii}) in systems 2, 3, 4 and 6, and
the ratio $N$(O\,{\sc vi}) to $N$(N\,{\sc v}) in the high-ionization phase by
varying the ionization parameter $\Gamma$. 
We assume that both gas zones are ionized by the Haardt \& Madau 
radiation field, and have the same
{\it relative} abundances as found in the DLA gas. Since the column density 
ratios are quite insensitive to the metallicity of the gas [M/H] for a
wide range of ionization parameters, we can constrain [M/H] using the
individual observed column densities.

Clearly, the assumption of same relative  abundances in A and B must not hold 
for gas-phase abundances of elements known to
be depleted by condensation into dust grains. If the dust-to-gas ratio
in the halo of 
this damped Ly$\alpha$ galaxy were considerably lower than in its DLA 
region (or even zero), photoionization models should arrive at ionic 
column densities of refractory elements that are underestimated, compared
with the observed value. We observe this effect for Fe\,{\sc ii} (see next 
section). 

\subsection{Low-ionization Phase}
\label{sec_low}

\subsubsection{The Lyman-limit System at $z=1.66184$ toward HE
1104--1805 B}

Table~\ref{tbl-4} displays the observed column densities of ions
associated with the LLS at $z=1.66184$. The low and high ion velocity
profiles are shown in Figures~\ref{fig9} and~\ref{fig11}, respectively
(right-hand panels). Besides the DLA system, this is the
second most metal-rich system. Prominent ions observed are: Mg\,{\sc
ii}, Al\,{\sc ii}, Al\,{\sc iii}, Si\,{\sc ii}, Si\,{\sc iii}, C\,{\sc
ii}, N\,{\sc iii} and Fe\,{\sc ii}. 
The detection of Fe\,{\sc ii} is marginal but
significant. Due to the low S/N and extraction artifacts at the
position of Fe\,{\sc ii} $\lambda 1608$ (Fig.~\ref{fig9}, top right
spectrum), we decided to fit one-component Voigt profiles to the three
strongest Fe\,{\sc ii} lines in the {\it NTT} spectra. The detection
of Al\,{\sc ii} and Al\,{\sc iii} is qualitatively consistent with the
assumption of DLA relative metal abundances in the LLS.

\placetable{tbl-4}

Taking the arguments given in \S~\ref{sec_geo} into account, 
in order to perform photoionization simulations for the LLS 
we have to first determine the range of possible values within which 
$N$(C\,{\sc iv})/$N$(Mg\,{\sc ii}) is likely to vary. 
Since the line profiles of Mg\,{\sc ii} do not trace the whole 
velocity range of C\,{\sc iv}, 
a very conservative upper limit of $\sim 20$ for the ratios is given by 
the {\it total} column densities of the fitted Voigt profiles. 
However, from the right-hand panel of Fig.~\ref{fig11} we see that the low ion
profiles (here represented by Al\,{\sc ii} $\lambda 1670$) coincide
quite well in velocity space with C\,{\sc iv} fit components 2 and
3. Considering that these Mg\,{\sc ii} line profiles at Keck HIRES
resolution would not differ too much from the Al\,{\sc ii} $\lambda
1670$  ones, and 
taking the column densities integrated over these components leads to 
$N$(C\,{\sc iv})/$N$(Mg\,{\sc ii})$=5.3$, a much more realistic column
density ratio. 

Table~\ref{tbl-6} displays the column densities
predicted by CLOUDY for the LLS using the Haardt \& Madau ionizing
radiation field (column labelled MODEL
1). Assuming for this system DLA gas-phase 
 abundances, a column density
ratio $N$(C\,{\sc iv})/$N$(Mg\,{\sc ii})$= 5.3$ implies 
$\log n_{\rm H}=-1.8$ cm$^{-3}$ 
(or $\log\Gamma=-2.95$ for this radiation field); in other
words, the gas is relatively highly ionized with 
$N$(H\,{\sc ii})/$N$(H\,{\sc i})$\approx 200$, but 
Mg\,{\sc ii} is still present. 
The observed log\,$N$(H\,{\sc i}) = 17.57 leads to
a typical (model dependent) cloud size along the LOSs of 
$S_{\|}\leq 1.6$ kpc. 
For comparison purposes, a power law of the form $f\propto\nu^{-2}$ was also 
used as ionizing background (column labeled MODEL 2 in
Table~\ref{tbl-6}). A harder radiation
field is not able to reproduce $N$(C\,{\sc iv})/$N$(Mg\,{\sc
ii})$= 5.3$ with the assumed element abundances. However, even the 
selected power law model fails to simultaneously reproduce 
C\,{\sc ii} and C\,{\sc iv}, while the Haardt \& Madau model does, due
to the continuum break at the He\,{\sc ii} edge.

\placetable{tbl-6}

Our model assumes that the LLS gas is in
photoionization equilibrium. Deviations from this equilibrium can lead to
dramatic underestimations of the cloud lengths parallel to the LOSs 
by photoionization
simulations, because the neutral hydrogen fraction is overestimated
(Haehnelt et al.~\cite{Haehnelt1}). However, at the density derived 
for this LLS, $n_{\rm H}\sim0.01$ cm$^{-3}$, significant departures
from the equilibrium temperature ($T=1.5\times10^{4}$ K) are rare, as is
shown by Smoothed Particle Hydrodynamics  simulations (Rauch et
al.~\cite{Rauch}, Haehnelt et 
al.~\cite{Haehnelt}). Thus, the hydrogen recombination timescale in this
regime, $\sim6$ Myr, is short enough to allow line cooling to balance
photoheating processes. Consequently, we believe the CLOUDY sizes
derived for this LLS to be reliable.

\paragraph{Gas Metallicity in the Lyman-limit System.}

The photoionization models described above are 
quite independent of the gas metallicity [M/H] over the
whole range of possible gas densities; therefore,
[M/H] can be determined by matching predicted  
and observed column densities. 
We find that, regardless of which model
is assumed, [M/H]$_{\rm LLS}={\rm [M/H]}_{\rm DLA}-0.2$ represents
the best prediction for nine ions observed in this system (Al\,{\sc
ii}, Al\,{\sc iii} and Fe\,{\sc ii} are not considered). 
In particular, $N$(C\,{\sc ii}), $N$(C\,{\sc iv}), $N$(Mg\,{\sc ii}), 
and $N$(Si\,{\sc iii}) are {\it simultaneously} very well reproduced if 
Z$_{\rm{LLS}}$ = 0.63~Z$_{\rm{DLA}}$.  
This is an upper limit for Z$_{\rm{LLS}}$ because due to saturation of the
Mg\,{\sc ii} lines in A, we obtain only a lower limit for [Mg/H];
hence, to reproduce $N$(Mg\,{\sc ii}) in B, 
a larger Mg content would require an even lower metallicity in
the LLS relative to the DLA gas. 
Varying the relative abundances within the observational errors in
the column densities leads us to estimate that this result is
significant at the 2$\sigma$ level (for comparison, considering solar
relative abundances in the LLS leads to [M/H]$_{\rm{LLS}}=-1.5$).

Studies of our galaxy halo gas show no systematic differences in the gas-phase 
abundances within galactocentric distances of 7 to 10 kpc in various 
directions, suggesting uniform physical properties over these radii. 
In addition, warm disk clouds show gas
abundances that are 0.2 to 0.6 dex lower than those in warm halo
clouds (Sembach \&{} Savage~\cite{Sembach}, and references therein; Savage \&{}
Sembach~\cite{Savage}, but see Cardelli et al.~\cite{Cardelli2}). 
If we assume that the LLS observed
at $z=1.66184$ in HE 1104--1805 B arises in the halo of the $z=1.66162$ 
damped Ly$\alpha$ galaxy seen in A, a negative gradient in metallicity
from DLA to halo gas 
implies that this halo gas has not yet been fully enriched with metals. 
This might be a consequence of different star-formation rates, in
which case LOS B would be probing gas regions of lower star-formation
rate than LOS A. Alternatively, such an abundance gradient could also
be explained by a disk-to-halo gas (and dust) transfer yet in early 
stages, if gas in the halo originates in the disk.

\paragraph{Fe and Al Abundances in the Lyman-limit System.}

Singly ionized iron is the most outstanding outlier in our model,
yielding an Fe\,{\sc ii} abundance one order of magnitude lower than 
observed (cf. Table~\ref{tbl-6}). We conclude that this difference can
only be due to different iron 
gas abundances between A and B, qualitatively consistent with the absence of
dust in the halo of this damped Ly$\alpha$ galaxy. This situation
resembles that in our Galaxy, where the degree of dust depletion
in halo clouds is smaller than in Disk clouds (Sembach \&{}
Savage~\cite{Sembach}). Further CLOUDY 
simulations show that the observed 
$\log N$(Fe\,{\sc ii})$=12.47$ can be reproduced if
[Fe/H]$_{\rm LLS}=-0.8$ (Haardt \& Madau~\cite{Haardt}) or $-1.0$ ($f_\nu\propto\nu^{-2}$). 
An opposite effect
is found for aluminum, where our model reproduces Al\,{\sc iii} only
if [Al/H]$=-1.0$, that is, if the Al overabundance is limited to the DLA 
gas (but such overabundance might be a consequence of a saturated
Al\,{\sc ii} $\lambda1670$ line in A [see also section~\ref{sec_lin}]).

\paragraph{A third ionization phase?}

$N$(C\,{\sc iv})/$N$(Mg\,{\sc ii})$=5.3$ also requires $N$(C\,{\sc iii})
to be lower by 1.5~$\sigma$ than observed. 
This suggests that there might in fact be a {\it
third} ``intermediate-ionization gas phase'', where {\it part of } 
the C\,{\sc iv}, C\,{\sc iii}, N\,{\sc iii} and Si\,{\sc iv} 
but no singly ionized species occur (neither does Si\,{\sc iii}, whose
ionizing potential of 33.5 eV is considerably lower than that of
C\,{\sc iii}). We would be thus observing blends
of lines arising in two phases at similar redshifts. The existence of
such a gas phase is fully consistent with the model
predictions for Si\,{\sc iii} and Al\,{\sc iii} in the low-ionization
phase (provided the Al
relative abundance is lower than in the DLA gas), and with the C\,{\sc iv}
line profiles, showing a much wider velocity span than the low ions
(Fig~\ref{fig11}). Unfortunately, the {\it HST} spectral resolution does not
allow an appropriate analysis of the  C\,{\sc iii} and N\,{\sc iii} 
line profiles.

\subsubsection{The {\rm C\,{\sc iv}} Systems at $z=1.66280$  
and $z=1.66465$ toward HE 1104--1805 A}

The $N$(C\,{\sc iv})/$N$(Mg\,{\sc ii}) ratios in the C\,{\sc iv} systems
at $z=1.66280$ and $z=1.66460$ in A (systems 2 and 3 in
Fig.~\ref{fig2}) are relatively well constrained by the observations,
so we basically repeated the procedure described in the previous
section, i.e., we searched for CLOUDY solutions that reproduce
these ratios by varying $\Gamma$ and $Z$. Since only the total neutral
hydrogen column density is known, $N$(H\,{\sc i}) was distributed
according to the $N$(Mg\,{\sc ii}) ratios. 

Table~\ref{tbl-8} displays the column densities
predicted by CLOUDY for System 3. 
Again assuming DLA gas-phase abundances, and the Haardt \& Madau 
metagalactic radiation
field, we find for this system that the
observed $N$(C\,{\sc iv})/$N$(Mg\,{\sc ii})$=70.8$ can be well
reproduced if $\log n_{\rm H}= -2.21$ cm$^{-3}$ 
(or $\log\Gamma=-2.54$ for this radiation field), 
implying a longitudinal cloud size of 
$S_{\|}\leq 0.9$ kpc, where the inequality stands for [Mg/H] $>-0.97$.
This size estimate assumes both C\,{\sc iv} and Mg\,{\sc ii} to arise 
from the same density region in photoionization equilibrium.

\placetable{tbl-7}

\placetable{tbl-8}

In System 2, the detection of Mg\,{\sc ii} is uncertain, and we can
only derive $N$(C\,{\sc iv})/$N$(Mg\,{\sc ii})$>117.5$, which
requires that $\log n_{\rm H}< -2.30$ ($\log\Gamma>-2.45$), or 
$S_{\|}$ \rlap{\lower 3.5 pt \hbox{$\mathchar \sim$}}\raise 1pt
\hbox {$>$} $0.5$ kpc for $N$(H\,{\sc
i})$=16.05$. Predicted column densities for system 2 are shown in
Table~\ref{tbl-7}. The disagreement for 
$N$(C\,{\sc iii}) arises as a consequence of a bad fit to the C\,{\sc
iii} $\lambda 977$ line.

None of these photoionization models necessarily requires that $Z<Z_{\rm DLA}$.

\subsubsection{The {\rm C\,{\sc iv}} Systems at $z=1.66493$ toward HE
1104--1805 B}

Observed and predicted column densities for the C\,{\sc iv} system 
at $z=1.66493$ in B (system 6) are shown in Table~\ref{tbl-9}. 
Because the detection of
Mg\,{\sc ii} is very uncertain, our model was required to
reproduce C\,{\sc iv} assuming $Z=0.63~Z_{\rm DLA}$. We arrived at $\log
n_{\rm H}= -2.40$ (or $\log\Gamma=-2.35$), implying $S_{\|}\sim 0.6$ kpc.

\placetable{tbl-9}

\subsection{{\rm O\,{\sc vi}}-phase}
\label{sec_high}

Possibly the most interesting result of this study is the significant 
detection of strong O\,{\sc vi} absorption at $z=1.66$ in both LOSs 
toward HE 1104--1805. Fig.~\ref{fig3} shows the corresponding 
section of the {\it HST} spectra of the A and B images with tickmarks above
the spectrum indicating the O\,{\sc vi} $\lambda\lambda$1031,1037
doublet at $z=1.66253$ and the C\,{\sc ii} $\lambda$1036 lines 
associated with the damped Ly$\alpha$ and Lyman-limit systems. 
Tickmarks below the flux level indicate lines
identified with other systems.
The O\,{\sc vi} $\lambda$1031 lines at $\lambda$ = 2747 \AA{} have 
similar rest-frame equivalent widths of $W_{\rm r}=0.93\pm0.07$ \AA{} 
(A) and $0.86\pm0.10$ \AA{}. 
Besides Si\,{\sc iii} $\lambda$1206 at $z=1.28$, neither
further metal lines nor a Ly$\beta$ line has been identified 
at this wavelength, but 
contamination by a weak Ly$\alpha$ line is not ruled out.\footnote{It seems
very unlikely that the absorption feature at $\lambda\sim2747$ \AA{} is 
{\it entirely} due to a Ly$\alpha$ forest interloper. In such a case,
two Ly$\alpha$ lines with $N({\rm H\,{\sc i}})\sim10^{16}$ cm$^{-2}$ ($b=30$
km~s$^{-1}$) would be required to reproduce the absorption profiles;
however, no associated C\,{\sc iv} is found, as it would be expected 
in this class of Ly$\alpha$ clouds.} 
Voigt profiles convolved with the instrumental profile 
have been overplotted in the spectra of A and B. They have Doppler parameters
$b=180$ km~s$^{-1}$ (A) and $110$ km~s$^{-1}$ (B) 
(obtained from Gaussian profile fits to the O\,{\sc vi} 
$\lambda$1031 lines) and common column density log\,$N=14.95$.
$N$ shows little variation with $b$ within
110--180 km~s$^{-1}$ in this region of the curve of growth. Clearly,
these large Doppler values do not necessarily represent the true line
widths. Rather, the observed line profiles are probably made up of
more than one O\,{\sc vi} line (but see next paragraphs). For comparison,
the dotted lines show the resulting profiles of 3 (A) and 2 (B) 
O\,{\sc vi} doublets with common $b$ = 50 km~s$^{-1}$ and {\it total} column
density $\log N=15.1$ (A) and $15.0$ (B). The doublets are placed at
redshifts corresponding to systems 1, 2, and 3 in A, and 4 and 5 in B
(cf. Fig.~\ref{fig2}). 

\placefigure{fig3}

Although O\,{\sc vi} has been shown to be relatively common in QSO
absorption systems (Bahcall et al.~\cite{Bahcall}; Bergeron et 
al.~\cite{Bergeron}; Burles \& Tytler~\cite{Burles}), 
its detection along the two LOSs to HE 1104--1805 at $z=1.66$ 
allows us to directly prove for the first time 
that this ion does indeed arise in very
extended gas clouds. There were indications of this for systems at lower
redshift (e.g. Bergeron et al.~\cite{Bergeron}; Lu \& Savage~\cite{Lu}) 
but no firm evidence. Large, low-density O\,{\sc vi} gas clouds
confirms the prediction by Rauch et al. (\cite{Rauch}) based on 
simulations of protogalactic clumps (PGCs). In this  model, PGCs with
a few times $10^9$ M$_\odot$ (in baryons) are embedded in a
filamentary, low density and largely
featureless O\,{\sc vi} phase with spatial extent of up to several 
hundreds kpc at the $N({\rm O\,{\sc vi}})=10^{13}$ cm$^{-2}$ column density
contour (cf. Fig. 3 in Rauch et al.). Detectable O\,{\sc vi} is indeed
expected to be more extended than C\,{\sc iv}. The gas is
photoionized, and the lines are bulk motion broadened and
wider than C\,{\sc iv}, as they sample the peculiar velocities over a
larger volume.

Observational evidence for the physical processes giving rise to O\,{\sc
vi}---collisional or photoionization---remain, however, 
debatable so far, basically because contamination by
Ly$\alpha$ lines makes it difficult to resolve 
the O\,{\sc vi} $\lambda\lambda1032,1036$ doublet profiles
appropriately. In our case, a quantitative assessment of 
the nature of this gas phase 
is possible, since no N\,{\sc v} is detected at this redshift 
(Fig.~\ref{fig4}), and photoionization models can be well 
constrained under certain assumptions.

\placefigure{fig4}

At the hydrogen density $n_{\rm H}\sim0.01$ cm$^{-3}$ 
derived for the C\,{\sc iv}+Mg\,{\sc ii} clouds, 
the fraction of ionizing photons with the 
$E>114$ eV necessary to ionize enough O\,{\sc v} into ``observable'' O\,{\sc vi}
is too small. Consequently, an additional, less dense phase is
required to explain the strong O\,{\sc vi} absorption observed in A
and B. Since a unique interpretation of the line profiles is
difficult with the present {\it HST} data, we made the following, simplifying
assumptions: (1) both LOSs pass through a common phase giving
rise to the O\,{\sc vi} absorption lines, suggested by the similar
equivalent widths and the zero velocity difference between lines in A
and B; and (2) the absorption lines arise in one cloud, suggested by the
line profile symmetry (this assumption may not be valid, but it is
equivalent to integrating  $N$(O\,{\sc vi}) over several line
components). A $3 \sigma$ upper limit of 
$W_{\rm r}=0.034$ \AA{} for the N\,{\sc v} $\lambda$1238.8210 line
(the stronger line of the N\,{\sc v} doublet) measured in the {\it
AAT} spectrum of A leads to a $3 \sigma$ upper limit of $\log
N$(N\,{\sc v}) $<13.2$. 
The non-detection of N\,{\sc v} places the stringent (but otherwise
very conservative) lower limit of 
$N$(O\,{\sc vi})/$N$(N\,{\sc v}) $>60$. This ratio requires densities
lower  than log\,$n_{\rm H}=-3.55$, 
or a large ionization parameter $\log\Gamma\geq -1.2$, 
if we assume the same relative abundances as found in the DLA gas.
This is a striking result, given
that a lower (closer to solar) [O/N] ratio would lead to
even less dense O\,{\sc vi} clouds.
In this regard, we must bear in mind that the CLOUDY 
simulations above depend partly 
on the relative abundances assumed. For instance, variations of the 
[O/N] ratio within the estimated column density uncertainties 
will lead to differences in the predicted parameters.

It is nontrivial to obtain an overall picture of this phase because
the total hydrogen column density is not known. As a consequence, 
photoionization models will not reproduce the individual 
observed column densities  $N$(O\,{\sc vi}) and $N$(N\,{\sc v}) uniquely, 
but both $N$(H) and the gas metallicity $Z$ will determine them. This
is shown in Fig.~\ref{fig5}, where we have plotted all the pairs
($Z$,$N$(H)) that
reproduce $\log N$(O\,{\sc vi}) $=15.0$ {\it and}  $\log N$(N\,{\sc
v}) $=13.2$. We can
see that, in general, a lower metallicity of the highly ionized gas 
relative to that of the DLA gas, say $Z<0.63Z_{\rm DLA}$, will be compatible 
with the observations if $N$(H) is large, $\log N$(H) $>20.0$. 
Additionally, we find that, regardless of the gas metallicity, the
clouds where this phase occurs must be at least one to two orders of
magnitude larger than the systems giving rise to Mg\,{\sc ii} 
absorption. These cloud sizes ($S_{\|}>\sim 100$ kpc), if
representative of transverse dimensions, are widely consistent with 
the detection of O\,{\sc vi} of similar strength in both LOSs. Typical
masses derived from these models are $\sim 10^9$ M$_\odot$.

\placefigure{fig5}

To see if C\,{\sc iv} is expected in such a model, 
we performed further ionization simulations for larger ionization
parameters, requiring that the assumed total hydrogen column density and 
metallicity always led to $\log N$(O\,{\sc vi}) $=15.0$, 
i.e., curves departing from 
points on the $Z$ vs. $N$(H) curve in Fig.~\ref{fig5}. The 
predicted $N$(C\,{\sc iv}) column densities in such models are shown
in Fig.~\ref{fig12} for the Haardt \& Madau (solid line) and a power-law
radiation 
field (dashed line). At the minimum 
value of $\Gamma$ allowed by the
observations $\log\Gamma=-1.2$, and for a wide range of gas
metallicities, CLOUDY predicts 
$\log N$(C\,{\sc iv})$\approx14.4$, in agreement with the observed
value. However, 
at higher ionization parameters, e.g.,  $N$(O\,{\sc vi})/$N$(N\,{\sc v})
$\sim 90$ or $\log\Gamma=-1.0$ 
(corresponding to a 2$\sigma$ significant non-detection of N\,{\sc v}), 
$N$(C\,{\sc iv}) becomes $\sim 40$\% smaller, independently of $Z$. 
This result in turn indicates that at least part of the observed C\,{\sc
iv} arises in highly ionized gas, with the remaining contribution coming from
the low-ionization gas phase. The present {\it HST} data do not 
enable us to quantitatively  assess this issue and we can only conclude that 
C\,{\sc iv} is likely to occur in both phases.

\placefigure{fig12}

Collisional ionization of gas in thermal equilibrium can also reproduce the
observed $N$(O\,{\sc vi})/$N$(N\,{\sc v}) ratio at 
$T=10^{5.45}$ K if one assumes solar
relative abundances (Sutherland \&{} Dopita \cite{Sutherland}), so
we cannot rule out this process as responsible for ionizing O\,{\sc v}
into O\,{\sc vi}. In such a case, the line broadening would be mostly
due to macro-turbulence. Collisionally ionized O\,{\sc vi} gas in a
Lyman-limit system at $z\simeq 3.4$ has been recently proposed by 
Kirkman \&{} Tytler (\cite{Kirkman}), who detect both C\,{\sc iv} and 
O\,{\sc vi} at the same redshift in high resolution data. These
authors favor collisional ionization due to the high kinetic
temperatures implied by the line widths of both ions. In
the case of HE 1104--1805 AB, additional, higher resolution ($ R \sim 23\,000
$) UV spectra are needed, in order 
to finally discriminate between collisional and photoionization as the
dominant ionization mechanism in the O\,{\sc vi}-phase.

\section{Conclusions and Final Remarks}
\label{sec_con}

We outline our conclusions as follows:

\begin{enumerate}
\item The damped Ly$\alpha$ system observed at $z=1.66162$ in the 
  ultraviolet and optical spectra of HE 1104--1805 A has a neutral
  hydrogen column density of $\log N$(H\,{\sc i})$=20.85\pm 0.01$. 
  The $6.6$ km~s$^{-1}$ resolution line profiles show that this
  high-density gas is distributed in four
  clouds spanning $\sim 80$ km~s$^{-1}$. Since the low ionization and
  Al\,{\sc iii}
  line profiles are observed to track fairly well, these ions
  are likely to arise in common clouds.
  Their  $b$-values are consistent 
  if the broadening mechanism is {\it not} purely thermal. 
  The C\,{\sc iv} line profiles show that at least some of C\,{\sc iv}
  absorption occurs in clouds with no singly ionized species.
  We find a metallicity of $Z_{\rm DLA}\simeq 1/10~Z_\odot$ 
  and a dust-to-gas ratio
  of $\sim 0.11$, based on [Zn/H]$=-1.02$ and [Cr/H]$=-1.46$. Other element 
  abundances are: [C/H]$=-0.91$, [N/H]$=-1.86$, [O/H]$=-0.98$, 
  [Mg/H]$=-0.97$, [Al/H]$=-0.77$, [Si/H]$=-1.02$, [Ti/H]$=-1.50$, 
  [Mn/H]$=-1.61$, [Fe/H]$=-1.59$ and [Ni/H]$=-1.68$. 
  The O/N ratio, 8 times larger than the solar value, is
  consistent with galactic chemical evolution models. However, the
  observed elemental abundance pattern must be modified by
  condensation into dust grains given (a) the underabundance of Cr and
  Ti relative to Zn; and (b) the 
  variation of the abundance ratios of these and other refractory
  elements among different clouds associated with the DLA system. 
\item Photoionization by the metagalactic radiation field proposed by Haardt
  \&{} Madau (\cite{Haardt}) 
  implies that the Lyman-limit System  observed at $z=1.66184$ 
  in the spectra of HE
  1104--1805 B and the three  further C\,{\sc iv} systems observed at
  similar  redshifts in A and B all belong to the same category of
  absorbers, namely small (LOS sizes 0.5 to 1.6 kpc) and relatively
  dense ($n_{\rm H}=0.01$ cm$^{-3}$) clouds where both C\,{\sc iv} and
  Mg\,{\sc ii} absorption occurs.
\item We find evidence for the LLS and one further C\,{\sc iv}
  absorption system observed in B to have lower metallicities 
  than observed in the DLA gas, $Z=0.63~Z_{\rm{DLA}}$. This result 
  is significant at the 2 $\sigma$ confidence level if the
  column density uncertainties are not underestimated.
\item In the context of galaxy formation, the observed H\,{\sc i}
  column densities suggest that LOS B
  intercepts gas in the ``halo'' of a  protogalaxy at $z$ =
  1.66, while LOS A crosses its denser, central gas regions. If this
  picture  is
  correct, then we are observing metal-poor gas in the halo of such a
  galaxy, compared to the high-density gas (within a transverse separation
  of $\sim$8~$h_{50}^{-1}$ kpc). This in turn suggests regions of different
  star-formation rates or, alternatively, metal enrichment through gas
  transfer from the inner to the outer regions of the protogalaxy
  still in early stages. 
\item The presence of a highly ionized gas phase giving rise to the
  observed O\,{\sc vi} absorption of similar strength in both spectra is
  evident. If photoionization is assumed, 
  the observed $\log N({\rm O\,{\sc vi}})=15.0$, and the non-detection
  of N\,{\sc v} implies large ($\sim100$--$200$  kpc),
  low-density ($\sim$10$^{-4}$ cm$^{-3}$) clouds where strong O\,{\sc vi}
  but weak C\,{\sc iv} absorption occurs. On the basis of these
  results, we suggest that this highly
  ionized gas surrounds the clouds giving rise to the damped
  Ly$\alpha$  and Lyman-limit systems. Such an extended low-density
  O\,{\sc vi} phase confirms the predictions of simulations of
  protogalactic clumps (Rauch et al.~\cite{Rauch}).
\end{enumerate}

Concerning the negative gradient in metallicity from gas crossed by
LOS A to gas crossed by LOS B, it must be pointed out that such an
effect, if real, does not necessarily support 
the disk/halo scenario. Indeed, it is still not clear whether
present-day rotating disk-like galaxies should have had a similar
appearance  in
their early stages of formation, nor is it well understood what we
should define as protogalactic ``halos''. A good deal of 
progress is being made to resolve this paradigm: a link between metallicity
and kinematics can improve our understanding of DLA systems (Wolfe et
al.~\cite{Wolfe2});  
hydrodynamic simulations of merging protogalactic clumps are capable
of explaining the large velocity spans seen in this class of
absorption systems (Haehnelt et al.~\cite{Haehnelt2}).
Here we have demonstrated the powerful tool which double LOSs can
provide us. A convincing interpretation of our result, however,  has
to await a larger sample of such cases.

Finally, let us emphasize that the scenario of small and compact
C\,{\sc iv} clouds surrounded by extended O\,{\sc vi} 
gas can indeed be theoretically understood and
is in good agreement with the predictions of a cold-dark-matter based
model (Rauch et al.~\cite{Rauch}). Furthermore, in a hierarchical
structure formation scenario damped Ly$\alpha$ and Lyman limit systems
can arise from relatively small (M$_{\rm baryon}\sim 10^{9}$
M$_\odot$) merging protogalactic clumps. Even if the sizes
derived using our photoionization models
are not representative of transverse cloud sizes, the similar 
C\,{\sc iv} line profiles and equivalent widths in A and B are still 
consistent with filamentary structures, which is also a prediction of 
hierarchical structure formation. Such a correlation between
line profiles in A and B is, on the other hand, less evident for
Mg\,{\sc ii}, implying gas inhomogeneities on spatial  scales similar
to the  separation between LOSs, and in concordance with Mg\,{\sc ii}
arising in smaller regions than C\,{\sc iv} and O\,{\sc vi}.

\acknowledgments

S. L. thanks the BMBF (DARA) for support under grant
No. 50OR96016; M. R. thanks NASA for support through grant HF-0107501-94A
from the Space Telescope Science Institute; 
W. L. W. S. was supported by grant AST 9529073 from the National Science
Foundation. A. S. thanks the financial
support under grant no. 
781-73-058 from the Netherlands Foundation for Research in Astronomy 
(ASTRON), which receives its funds from the Netherlands Organisation
for Scientific Research (NWO). We thank the anonymous referee for many 
helpful suggestions.

\clearpage
 
\begin{deluxetable}{llcccc}
\tablecaption{Journal of Observations. \label{tbl-5}}
\tablewidth{0pt}
\tablehead{
\colhead{Object}&\colhead{ID\tablenotemark{a}}&\colhead{UT date}
&\colhead{Coverage
(\AA)}&\colhead{S/N\tablenotemark{b}}&\colhead{FWHM (\AA)}
}
\startdata
HE 1104--1805 A & {\it HST} spectra & Nov. 1995 & 2222--3277 & 20 & 2.0\nl
HE 1104--1805 B &&&&17&\nl
HE 1104--1805 A&{\it AAT} spectra\tablenotemark{c}&
May 1993&3170--7570&82&1.2\nl
HE 1104--1805 B&&&&30&\nl
HE 1104--1805 A&{\it {\it NTT}} spectra&Feb. 1996&3910--8290&77&0.8--1.0\nl
HE 1104--1805 B&&&&27&\nl
HE 1104--1805 A&Keck spectra&Jan. and Feb. 1997&3620--6080&61&0.08--0.13\nl
HE 1104--1805 B&&&&42&\nl
\enddata
\tablenotetext{a}{{\it HST}: Hubble Space Telescope; {\it AAT}:
Anglo-Australian Telescope; {\it {\it NTT}}: New-Technology Telescope.}
\tablenotetext{b}{Maximum signal-to-noise ratio per pixel.}
\tablenotetext{c}{Reference: Paper I}
\end{deluxetable}

\clearpage
 
\begin{deluxetable}{ccccc}
\tablecaption{Lyman-edges in HE 1104-1805 A and B. \label{tbl-1}}
\tablewidth{0pt}
\tablehead{
\colhead{} & \colhead{$z_{\rm LLS}$}  &\colhead{$\tau$\tablenotemark{a}}
& \colhead{log\,$N_{\rm{H\,{\sc i}}}$(cm$^{-2}$)}
& \colhead{$b$\,(km~s$^{-1}$)} 
}
\startdata
A & 1.66162 &\nodata      & 20.84\tablenotemark{b}  & 27 \nl
  & 2.20016 &0.29$\pm$0.08& 16.67$\pm$0.12& 35 \nl
  & 2.29880 &0.14$\pm$0.07& 16.36$\pm$0.23& 40 \nl
\tablevspace{0.5cm}
B & 1.66184 &2.34$\pm$0.51& 17.57$\pm$0.10& 30 \nl
  & 2.20099 &0.63$\pm$0.11& 17.00$\pm$0.08& 40 \nl
  & 2.29877 &0.21$\pm$0.11& 16.53$\pm$0.23 & 35 \nl
\enddata
\tablenotetext{a}{Optical depth $\tau=\ln(f_{+}/f_{-})$, where
$f_{+}$ is the extrapolated continuum redward of $912\times(1+z_{\rm
LLS})$, and  $f_{-}$ the absorbed continuum.}        
\tablenotetext{b}{Column density constrained by the damped          
Ly$\alpha$ wings.}
\end{deluxetable}

\clearpage
 
\begin{deluxetable}{lrlcclclc}                              
\scriptsize                                                          
\tablecaption{Line Parameters for $z_{\rm abs}=1.66$ 
Absorption Systems toward HE 1104--1805 A. \label{tbl-2}}              
\tablewidth{0pt}                                                              
\tablehead{                                                                   
\colhead{Species} & \colhead{Lines}&\colhead{$z$}
&\colhead{$W_{\rm r}$(\AA)\tablenotemark{a}}&
\colhead{$b$\,(km~s$^{-1}$)}&
\colhead{$\sigma_{b}$\tablenotemark{b}}
&\colhead{log\,$N$(cm$^{-2}$)}
&\colhead{$\sigma_{{\rm log}N}$}&\colhead{Spectrum\tablenotemark{c}} 
}                                                                             
\startdata                                                                    
H\,{\sc i} &1215  &  1.661620  & 19.05  & 26.90   & 0.30 & 20.85  &0.01 &4 \nl 
    &1025  &          	&  2.99  &         &     &        &     &4 \nl
    &$\vdots$&          &\nodata &         &     &        &     &4 \nl
    &918   &          	&\nodata &         &     &        &     &4 \nl
    &1215  &  1.664490	&        & 28.50   &0.90 & 16.59  &0.07 &4 \nl  
    &1025  &         	&        &         &     &        &     &4 \nl
    &$\vdots$&         	&        &         &     &        &     &4 \nl
    &918   &          	&        &         &     &        &     &4 \nl  
C\,{\sc i}  &1656  &  1.661620   &$<$0.02 & 20.00   & 0.00    &$<$12.80&\nodata&3 \nl  
C\,{\sc ii} &1036  &  1.661640	&\nodata & 20.00   & 0.00    & 16.50  &0.20 &4 \nl  
    &1334  &     	&  0.46  &       &     &      &     &3 \nl  
C\,{\sc iii}&977   &  1.661430	&  1.62  & 44.00   & 0.00    & 15.68  &0.20 &4 \nl  
            &977   &  1.662800	&        & 13.10   &0.00     & 13.15  &     &4 \nl  
            &977   &  1.664600	&        & 15.80   &0.00     & 15.42  &     &4 \nl	
C\,{\sc iv} &1548  &  1.661186	&  0.361 &  15.88  &1.89 &13.31   &0.16 &1 \nl
    &1550  &        	&  0.197 &         &     &        &     &1 \nl
    &1548  &  1.661412	&        &  22.44  &3.14 &13.87   &0.06 &1 \nl
    &1550  &        	&        &         &     &        &     &1 \nl
    &1548  &  1.661694	&        &  13.29  &2.94 &13.28   &0.21 &1 \nl
    &1550  &        	&        &         &     &        &     &1 \nl
    &1548  &  1.661908	&        &  28.51  &5.09 &13.34   &0.13 &1 \nl
    &1550  &        	&        &         &     &        &     &1 \nl
    &1548  &  1.662776	&  0.147 &  15.11  &4.09 &13.13   &0.33 &1 \nl
    &1550  &        	&  0.098 &         &     &        &     &1 \nl
    &1548  &  1.662856	&        &   8.82  &0.69 &13.73   &0.07 &1 \nl
    &1550  &        	&        &         &     &        &     &1 \nl
    &1548  &  1.662984	&        &  20.25  &12.40&12.88   &0.55 &1 \nl
    &1550  &        	&        &         &     &        &     &1 \nl
    &1548  &  1.664327	&  0.214 &   4.69  &1.81 &12.32   &0.10 &1 \nl
    &1550  &        	&  0.147 &         &     &        &     &1 \nl
    &1548  &  1.664598	&        &  13.42  &0.94 &13.86   &0.03 &1 \nl
    &1550  &        	&        &         &     &        &     &1 \nl
    &1548  &  1.664739	&        &   4.81  &0.97 &13.38   &0.08 &1 \nl
    &1550  &        	&        &         &     &        &     &1 \nl
    &1548  &  1.664860	&        &   4.52  &0.68 &13.07   &0.03 &1 \nl
    &1550  &        	&        &         &     &        &     &1 \nl
N\,{\sc i}  &1199  &  1.661640	&  0.29  & 20.00   &0.00     & 14.00  &0.20 &4 \nl  
            &1200.2&    	&        &         &         &&     &4 \nl  
            &1200.7&    	&        &         &         &        &     &4 \nl  
N\,{\sc ii} &1083 &  1.661640	&  0.53  & 20.00   &0.00     & 15.00  &0.20 &4 \nl  
N\,{\sc iii}&989   &  1.661430	&  0.44  & 44.00   &0.00     & 13.60  &0.20 &4 \nl  
N\,{\sc v}  &1238  &  1.662530	&$<$0.034&\nodata  &\nodata  &$<$13.20&\nodata&4 \nl  
O\,{\sc i} &988   &  1.661640	& \nodata& 20.00   &0.00     & 16.80  &0.20 &4 \nl  
    &1302  &       	&  0.75  &         &         &       &     &3 \nl  
O\,{\sc vi} &1031  &  1.662530	&  0.93  &180.00   &0.00     & 14.95  &0.20 &4 \nl  
    &1037  &     	&\nodata &       &     &       &     &4 \nl  
Mg\,{\sc i} &2852  &  1.661523	&  0.25  & 14.20   &0.00     & 12.33  &0.03 &2 \nl  
    &2852  &  1.661793	&        & 24.30   &     & 11.65  &0.09 &2 \nl  
Mg\,{\sc ii}&2796  &  1.661370	& 0.96   &   3.61  &0.00     &10.44   &0.06 &2 \nl
    &2803  &        	& 0.90   &         &     &        &     &2 \nl
    &2796  &  1.661549	&        &  16.69  &0.00     &15.09   &0.03 &2 \nl
    &2803  &        	&        &         &     &        &     &2 \nl
    &2796  &  1.661710	&        &   7.31  &0.00     &15.21   &0.18 &2 \nl
    &2803  &        	&        &         &     &        &     &2 \nl
    &2796  &  1.661982	&        &   8.70  &0.00     &13.30   &0.02 &2 \nl
    &2803  &        	&        &         &     &        &     &2 \nl
    &2796  &  1.662760	&$<$0.03 & 10.00   &0.00     &$<$11.80&\nodata     &2 \nl
    &2803  &        	&        &         &     &        &     &2 \nl  
    &2796  &  1.664650	&  0.06  &  7.70   &0.00     & 12.19  &0.03 &2 \nl  
    &2803  &    	&        &         &     &        &     &2 \nl  
Al\,{\sc i} &1765  &  1.661620	&$<$0.02 & 20.00   &\nodata&$<$12.10&\nodata     &1 \nl  
Al\,{\sc ii}&1670  &  1.661382	&0.404   &   3.21  &0.47 &14.66   &0.31 &1 \nl
    &1670  &  1.661485	&        &  12.29  &3.33 &12.97   &0.17 &1 \nl
    &1670  &  1.661698	&        &  12.77  &0.92 &13.12   &0.06 &1 \nl
    &1670  &  1.661990	&        &   7.18  &0.51 &12.21   &0.02 &1 \nl
Al\,{\sc iii}&1854 &  1.661378	& 0.153  &   2.99  &0.42 &12.34   &0.02 &1 \nl
     &1862 &        	& 0.092  &         &     &        &     &1 \nl
     &1854 &  1.661531	&        &  19.79  &1.51 &12.85   &0.03 &1 \nl
     &1862 &        	&        &         &     &        &     &1 \nl
     &1854 &  1.661727	&        &   7.34  &1.04 &12.27   &0.09 &1 \nl
     &1862 &        	&        &         &     &        &     &1 \nl
     &1854 &  1.661981	&        &   5.32  &1.93 &11.58   &0.08 &1 \nl
     &1862 &        	&        &         &     &        &     &1 \nl
Si\,{\sc i} &1845  &  1.661620	&$<$0.01 & 20.00   &\nodata&$<$12.90&\nodata     &1 \nl  
Si\,{\sc ii}&1808  &  1.661371	& 0.103  &   3.61  &0.19 &15.13   &0.02 &1 \nl
    &1526  &        	& 0.430  &         &     &        &     &1 \nl
    &1808  &  1.661538	&        &  16.69  &0.39 &14.97   &0.01 &1 \nl
    &1526  &        	&        &         &     &        &     &1 \nl
    &1808  &  1.661773	&        &   7.31  &1.08 &13.96   &0.09 &1 \nl
    &1526  &        	&        &         &     &        &     &1 \nl
    &1808  &  1.661984	&        &   8.70  &0.59 &13.66   &0.02 &1 \nl
    &1526  &        	&        &         &     &        &     &1 \nl
Si\,{\sc iii}&1206 &  1.661430	&  0.60  & 44.00   &0.00     & 14.00  &0.20 &4 \nl  
             &1206 &  1.664600	&        & 15.80   &0.00     & 13.70  &0.20     &4 \nl  
Si\,{\sc iv}&1393  &  1.661640	&$<$0.69  &\nodata&\nodata&$<$14.08&\nodata &1 \nl  
    &1402  &    	&$<$0.34  &         &     &        &     &1 \nl
    &1393  &  1.662800	&\nodata &\nodata &\nodata &$<$12.58&\nodata &1 \nl  
    &1402  &    	&$<$0.017&         &     &        &     &1 \nl
    &1393  &  1.664650	&0.170   &\nodata&\nodata  &13.11   &0.07 &1 \nl  
    &1402  &    	& 0.044  &         &     &        &     &1 \nl  
Ti\,{\sc ii}&1910.6&  1.661389	& 0.007  &   2.81  &1.16 &12.28   &0.05 &1 \nl
    &1910.9&        	& 0.003  &         &     &        &     &1 \nl
Cr\,{\sc ii}&2056  &  1.661381  & 0.051  &   3.94  &0.22 &12.86   &0.01 &1 \nl
    &2062  &    	& 0.043  &         &     &        &     &1 \nl
    &2066  &  	        & 0.028  &         &     &        &     &1 \nl
    &2056  &  1.661575	&        &   8.26  &0.73 &12.61   &0.02 &1 \nl
    &2062  &           	&        &         &     &        &     &1 \nl
    &2066  &            &        &         &     &        &     &1 \nl
    &2056  &  1.661705	&        &   1.75  &0.74 &11.68   &0.12 &1 \nl
    &2062  &            &        &         &     &        &     &1 \nl
    &2066  &            &        &         &     &        &     &1 \nl
Mn\,{\sc i}&2795   &  1.661420	&  0.06  & 14.20   &0.00 & 11.80  &0.20 &2 \nl  
Mn\,{\sc ii}&2576  &  1.661469	&  0.05  & 14.20   &0.00 & 12.66  &0.06 &2 \nl  
    &2594  &            &\nodata &         &     &        &     &2 \nl
    &2606  &            &\nodata &         &     &        &     &2 \nl
    &2576  &  1.661734	&        & 24.30   &     & 11.82  &0.38 &2 \nl  
    &2594  &            &        &         &     &        &     &2 \nl
    &2606  &            &        &         &     &        &     &2 \nl
Fe\,{\sc i}&2484   &  1.661620	&$<$ 0.02& 20.00 &\nodata&$<$12.30&\nodata&2 \nl  
Fe\,{\sc ii}&1608  &  1.661376	& 0.280  &   4.80  &0.16 &14.49   &0.01 &1 \nl
    &2249  &        	& 0.045  &         &     &        &     &1 \nl
    &2260  &        	& 0.079  &         &     &        &     &1 \nl
    &1608  &  1.661549	&        &   8.49  &0.70 &14.29   &0.03 &1 \nl
    &2249  &        	&        &         &     &        &     &1 \nl
    &2260  &        	&        &         &     &        &     &1 \nl
    &1608  &  1.661709	&        &   8.83  &1.11 &13.85   &0.06 &1 \nl
    &2249  &        	&        &         &     &        &     &1 \nl
    &2260  &        	&        &         &     &        &     &1 \nl
    &1608  &  1.661983	&        &   9.12  &1.10 &13.34   &0.04 &1 \nl
    &2249  &        	&        &         &     &        &     &1 \nl
    &2260  &        	&        &         &     &        &     &1 \nl
Ni\,{\sc ii}&1709  &  1.661382	& 0.052  &   3.05  &0.35 &13.07   &0.02 &1 \nl
    &1741  &        	& 0.051  &         &     &        &     &1 \nl
    &1751  &        	& 0.048  &         &     &        &     &1 \nl
    &1709  &  1.661559	&        &  11.52  &1.38 &12.98   &0.04 &1 \nl
    &1741  &        	&        &         &     &        &     &1 \nl
    &1751  &        	&        &         &     &        &     &1 \nl
    &1709  &  1.661716	&        &   0.52  &0.19 &12.29   &0.27 &1 \nl
    &1741  &        	&        &         &     &        &     &1 \nl
    &1751  &        	&        &         &     &        &     &1 \nl
    &1709  &  1.661921	&        &  21.67  &7.71 &12.47   &0.12 &1 \nl
    &1741  &        	&        &         &     &        &     &1 \nl
    &1751  &        	&        &         &     &        &     &1 \nl
Zn\,{\sc ii}&2062  &  1.661373	& 0.026  &   3.04  &0.24 &12.37   &0.01 &1 \nl
    &2026  &            & 0.042  &         &     &        &     &1 \nl
    &2062  &  1.661560	&        &   7.64  &1.21 &11.82   &0.04 &1 \nl
    &2026  &        	&        &         &     &        &     &1 \nl
\enddata					    
\tablenotetext{a}{Rest-frame equivalent widths. For blends, the total
equivalent width is given. Upper limits ($3\sigma$) 
represent non-detections.}		    
\tablenotetext{b}{$\sigma_{b}=0$ indicates fixed $b$ parameter.}
\tablenotetext{c}{(1) Keck; (2) {\it NTT}; (3) {\it AAT}; (4) {\it HST}}
\end{deluxetable}				    
						    
\clearpage						    

\begin{deluxetable}{lclclclclcl}            
\scriptsize                   
\tablecaption{Ionic Column Densities (cm$^{-2}$) and Gas-Phase 
Abundances in the $z_{\rm abs}=1.66162$ DLA 
toward HE 1104--1805 A. \label{tbl-3}}                           
\tablewidth{0pt}                                                              
\tablehead{                                                                   
\colhead{Species}
&\colhead{log\,$N_{\rm fit}(X_i)$}&\colhead{$\sigma_{{\rm log}N}$}   
&\colhead{log\,$N_{\rm app}(X_i)$}&\colhead{$\sigma_{{\rm log}N}$}   
&\colhead{log\,$N_{\rm ad}(X_i)$}&\colhead{$\sigma_{{\rm log}N}$}   
&\colhead{log\,$N(X)$}&\colhead{$\sigma_{{\rm log}N}$}   
&\colhead{$[X/{\rm H}]$\tablenotemark{a}}&\colhead{$\sigma_{[X/{\rm H}]}$}
}                                                                             
\startdata 
 H\,{\sc i}    &  20.85 & 0.01 &\nodata&\nodata&  20.85 & 0.01 &      &     &      &      \nl
 C\,{\sc i}    &$<$12.80&\nodata&\nodata&\nodata&$<$12.80&\nodata& 16.50& 0.20& -0.91& 0.20 \nl
 C\,{\sc ii}   &  16.50 & 0.20 &\nodata&\nodata&  16.50 & 0.20 &      &	    &      &	  \nl
 C\,{\sc iii}  &  15.70 & 0.20 &\nodata&\nodata&  15.70 & 0.20 &      &	    &      &	  \nl
 C\,{\sc iv}   &  14.13 & 0.05 & 14.13 &  0.01 &  14.13 & 0.05 &      &	    &      &	  \nl
 N\,{\sc i}    &  14.00 & 0.20 &\nodata&\nodata&  14.00 & 0.20 & 15.04& 0.16& -1.86& 0.16 \nl
 N\,{\sc ii}   &  15.00 & 0.20 &\nodata&\nodata&  15.00 & 0.20 &      &	    &      &	  \nl
 N\,{\sc iii}  &  13.60 & 0.20 &\nodata&\nodata&  13.60 & 0.20 &      &	    &      &	  \nl
 N\,{\sc v}    &$<$13.20&\nodata&\nodata&\nodata&$<$13.20&\nodata&    &	    &      &	  \nl
 O\,{\sc i}    &  16.80 & 0.20 &\nodata&\nodata&  16.80 & 0.20 & 16.80& 0.20& -0.98& 0.20 \nl
 O\,{\sc vi}   &  14.95 & 0.20 &\nodata&\nodata&  14.95 & 0.20 &      &	    &      &	  \nl
 Mg\,{\sc i}   &  12.41 & 0.09 &\nodata&\nodata&  12.41 & 0.09 &$>$15.46&0.09&$>$-0.97& 0.09 \nl
 Mg\,{\sc ii}  &  15.46 & 0.09 &\nodata&\nodata&$>$15.46& 0.09 &      &	    &      &	  \nl
 Al\,{\sc i}   &$<$12.10&\nodata&\nodata&\nodata&$<$12.10&\nodata& 14.56& 0.15& -0.77& 0.15 \nl
 Al\,{\sc ii}  &  14.55 & 0.24 & 13.43 &  0.16 &  14.55 & 0.24 &      &	    &      &	  \nl
 Al\,{\sc iii} &  13.06 & 0.02 & 13.08 &  0.01 &  13.06 & 0.02 &      &	    &      &	  \nl
 Si\,{\sc i}   &$<$12.90&\nodata&\nodata&\nodata&$<$12.90&\nodata& 15.38& 0.02& -1.02& 0.02 \nl
 Si\,{\sc ii}  &  15.38 & 0.02 & 15.26 &  0.01 &  15.38 & 0.02 &      &	    &      &	  \nl
 Si\,{\sc iii} &  14.00 & 0.20 &\nodata&\nodata&  14.00 & 0.20 &      &	    &      &	  \nl
 Si\,{\sc iv}  &\nodata &\nodata&$<$14.08&\nodata&$<$14.08&\nodata&   &	    &      &	  \nl
 Ti\,{\sc ii}  &  12.28 & 0.05 & 12.28 &  0.07 &  12.28 & 0.07 & 12.28& 0.07& -1.50& 0.07 \nl
 Cr\,{\sc ii}  &  13.07 & 0.01 & 13.05 &  0.01 &  13.07 & 0.01 & 13.07& 0.01& -1.46& 0.01 \nl
 Mn\,{\sc i}   &  11.80 & 0.20 &\nodata&\nodata&  11.80 & 0.20 & 12.77& 0.14& -1.61& 0.14 \nl
 Mn\,{\sc ii}  &  12.72 & 0.18 &\nodata&\nodata&  12.72 & 0.18 &      &	    &      &	  \nl
 Fe\,{\sc i}   &$<$12.30&\nodata&\nodata&\nodata&$<$12.30&\nodata& 14.77& 0.01& -1.59& 0.02 \nl
 Fe\,{\sc ii}  &  14.77 & 0.01 & 14.68 &  0.02 &  14.77 & 0.02 &      &	    &      &	  \nl
 Ni\,{\sc ii}  &  13.42 & 0.03 & 13.43 &  0.02 &  13.42 & 0.03 & 13.42& 0.03& -1.68& 0.03 \nl
 Zn\,{\sc ii}  &  12.48 & 0.01 & 12.45 &  0.01 &  12.48 & 0.01 & 12.48& 0.01& -1.02& 0.01 \nl
\enddata					    
\tablenotetext{a}{Relative to solar element abundances compiled by
Verner et al.~\cite{Verner}.}
\end{deluxetable}				    
 						    
\clearpage
 
\begin{deluxetable}{lrlcclclc}                              
\scriptsize                         
\tablecaption{Line Parameters for the $z_{\rm abs}=1.66$  
Absorption Systems toward HE 1104--1805 B. \label{tbl-4}}                  
\tablewidth{0pt}                                                              
\tablehead{                                                                   
\colhead{Species} & \colhead{Lines}&\colhead{$z$}&
\colhead{$W_{\rm r}$ (\AA)\tablenotemark{a}}&
\colhead{$b$\,(km~s$^{-1}$)}&
\colhead{$\sigma_{b}$\tablenotemark{b}}
&\colhead{log\,$N$(cm$^{-2}$)}
&\colhead{$\sigma_{{\rm log}N}$}&\colhead{Spectrum\tablenotemark{c}}       
}                                                                             
\startdata                                                                    
H\,{\sc i}  &1215   & 1.661840 & 1.690 &30.00 &0.00  & 17.57& 0.00 &4 \nl
    &1025   &          & 0.790 &      &     &        &      &4 \nl
    &$\vdots$&         &\nodata&      &     &        &      &4 \nl
    &918    &          &\nodata&      &     &        &      &4 \nl
    &1215   & 1.664970 &       &30.00 &0.00 & 15.90  & 0.13 &4 \nl
    &1025   &          &       &      &     &        &      &4 \nl
    &$\vdots$&         &       &      &     &        &      &4 \nl
    &918    &          &       &      &     &        &      &4 \nl
C\,{\sc i}  &1656   & 1.661650 &$<$0.09&20.00 &0.00 &$<$13.50&      &3 \nl
C\,{\sc ii} &1036   & 1.661650 &\nodata &20.00 &0.00 & 14.50  & 0.20 &4 \nl
    &1334   &          &  0.18&      &     &        &      &3 \nl
C\,{\sc iii}&977    & 1.664480 &  0.95&37.00 &0.00 & 15.50  & 0.20 &4 \nl
            &977    & 1.661830 &       &29.30 &0.00 & 14.00  & 0.20 &4 \nl
C\,{\sc iv}  &1548  & 1.661130 & 0.630 &19.93 &2.33 &  13.24 & 0.04 &1 \nl
     &1550  &          & 0.361 &      &     &        &      &1 \nl
     &1548  & 1.661512 &       & 8.82 &0.88 &  13.56 & 0.06 &1 \nl
     &1550  &          &       &      &     &        &      &1 \nl
     &1548  & 1.661759 &       &21.96 &4.49 &  13.81 & 0.10 &1 \nl  
     &1550  &          &       &      &     &        &      &1 \nl
     &1548  & 1.662036 &       &12.62 &1.74 &  13.77 & 0.09 &1 \nl
     &1550  &          &       &      &     &        &      &1 \nl
     &1548  & 1.662250 &       &47.39 &4.03 &  13.71 & 0.15 &1 \nl
     &1550  &          &       &      &     &        &      &1 \nl
     &1548  & 1.662878 &       &20.75 &3.44 &  13.31 & 0.10 &1 \nl
     &1550  &          &       &      &     &        &      &1 \nl
     &1548  & 1.663167 &       & 5.89 &2.53 &  12.38 & 0.16 &1 \nl
     &1550  &          &       &      &     &        &      &1 \nl
     &1548  & 1.663384 &       & 8.02 &2.11 &  12.55 & 0.08 &1 \nl
     &1550  &          &       &      &     &        &      &1 \nl
     &1548  & 1.664721 & 0.077 & 3.54 &1.49 &  12.61 & 0.07 &1 \nl
     &1550  &          & 0.067 &      &     &        &      &1 \nl
     &1548  & 1.664887 &       & 5.43 &0.36 &  13.63 & 0.03 &1 \nl
     &1550  &          &       &      &     &        &      &1 \nl
N\,{\sc i}  &1199   & 1.661650 &$<$0.11&20.00 &\nodata&$<$14.00&\nodata&4 \nl
N\,{\sc ii} &1083   & 1.661650 &$<$0.13&20.00 &\nodata&$<$13.70&\nodata&4 \nl
N\,{\sc iii}&989    & 1.661830 &0.27&37.00 &0.00     & 14.00  & 0.20 &4 \nl
    &989    & 1.662840 &       &29.90 &0.00     & 13.70  & 0.20 &4 \nl
    &989    & 1.665070 &   0.27&37.20 &0.00     & 14.20  & 0.20 &4 \nl
N\,{\sc v}  &1238   & 1.662530 &$<$0.19&     &\nodata&$<$14.00&\nodata&4 \nl
O\,{\sc i}  &988    & 1.661650 &$<$0.13&20.00 &\nodata&$<$14.10&\nodata&4 \nl
    &1302   &          &$<$0.19&      &     &       &      &3 \nl
O\,{\sc vi} &1031   & 1.662530 &0.86   &110.00&0.00 & 14.95  & 0.20 &4 \nl
    &1037   &          &       &      &     &        &      &4 \nl
Mg\,{\sc i} &2853   &1.661650  &$<$0.10&20.00 &\nodata&$<$11.90&\nodata&2 \nl
Mg\,{\sc ii} &2796  & 1.661557 & 0.343 & 7.83 &0.00 &  13.14 & 0.13 &2 \nl
     &2803  &          & 0.239 &      &     &        &      &2 \nl
     &2796  & 1.661783 &       &26.52 &0.00 &  12.73 & 0.09 &2 \nl
     &2803  &          &       &      &     &        &      &2 \nl
     &2796  & 1.664930 & $<$0.09&10.00 &     &$<$12.40&      &2 \nl
     &2803  &          &       &      &     &        &      &2 \nl
Al\,{\sc i}  &1765  &1.661650  &$<$0.10&20.00 &\nodata&$<$12.60&\nodata&1 \nl
Al\,{\sc ii} &1670  & 1.661556 & 0.097 & 7.70 & 0.00&  12.26 & 0.03 &1 \nl
             &1670  & 1.661781 &       &14.10 & 0.00&  11.80 & 0.06 &1 \nl
             &1670  & 1.661990 &       & 8.10 & 0.00&  11.45 & 0.10 &1 \nl
Al\,{\sc iii}&1854  & 1.661564 & 0.041 & 7.75 & 1.29&  12.08 & 0.11 &1 \nl
     &1862  &          & 0.036 &      &     &        &      &1 \nl
     &1854  & 1.661784 &       &14.07 &13.05&  11.98 & 0.35 &1 \nl
     &1862  &          &       &      &     &        &      &1 \nl
     &1854  & 1.661956 &       & 8.15 &11.22&  11.81 & 0.46 &1 \nl
     &1862  &          &       &      &     &        &      &1 \nl
Si\,{\sc i} &1845   & 1.661650 &$<$0.08&20.00 & \nodata&$<$13.20&\nodata&1 \nl
Si\,{\sc ii}&1526   & 1.661552 & 0.076 & 8.30 & 0.94&  13.29 & 0.03 &1 \nl
    &1526   & 1.661747 &       & 7.72 & 3.94&  12.80 & 0.15 &1 \nl
    &1526   & 1.661941 &       & 8.99 & 3.69&  12.89 & 0.12 &1 \nl
Si\,{\sc iii}&1206  & 1.661830 &0.55&37.00 &     & 14.10  & 0.20 &4 \nl
Si\,{\sc iv}&1393   & 1.661840 &1.166  &\nodata &\nodata&$<$14.28&\nodata &1 \nl
            &1402   &          &  0.580&      &     &        &      &1 \nl
            &1393   & 1.664930 &\nodata&\nodata &\nodata&$<$12.86&\nodata&1 \nl
            &1402   &          &$<$0.033 &      &     &        &      &1 \nl         
Fe\,{\sc ii}&2344   & 1.661578 & 0.012 & 7.86 & 4.30 &  12.47 & 0.08 &2 \nl
            &2382   &          & 0.030 &      &      &        &      &2 \nl
            &2600   &          & 0.045 &      &     &        &      &2 \nl

\enddata	             				    
\tablenotetext{a}{Rest-frame equivalent widths. For blends, the total
equivalent width is given. Upper limits ($3\sigma$) 
represent non-detections.}		    
\tablenotetext{b}{$\sigma_{b}=0$ indicates fixed $b$ parameter.}
\tablenotetext{c}{(1) Keck; (2) {\it NTT}; (3) {\it AAT}; (4) {\it HST}}
\end{deluxetable}				    
						    
\clearpage

\begin{deluxetable}{lrlclclcl}                                                
\scriptsize       
\tablecaption{Ionic Column Densities (cm$^{-2}$) in the LLS at $z_{\rm
abs}=1.66184$ toward HE 1104--1805 B in Comparison to Results from 
Photoionization Models. \label{tbl-6}}       
\tablewidth{0pt}                                                              
\tablehead{                                                                   
\colhead{Species}
&\colhead{log\,$N_{\rm fit}(X_i)$}&\colhead{$\sigma_{{\rm log}N}$}   
&\colhead{log\,$N_{\rm app}(X_i)$}&\colhead{$\sigma_{{\rm log}N}$}   
&\colhead{log\,$N_{\rm ad}(X_i)$}&\colhead{$\sigma_{{\rm log}N}$}   
&\colhead{MODEL 1\tablenotemark{a}}&\colhead{MODEL 2\tablenotemark{b}}
}                                                                             
\startdata 
 H\,{\sc i}    &  17.57 & 0.10  &\nodata&\nodata&  17.57 & 0.10 &17.57 &17.57 \nl
 H\,{\sc ii}   &\nodata &\nodata&\nodata&\nodata&\nodata &\nodata      &19.89 &19.82 \nl
 C\,{\sc i}    &$<$13.50&\nodata&\nodata&\nodata&$<$13.50&\nodata      &11.87 &12.14 \nl
 C\,{\sc ii}   &  14.50 & 0.20  &\nodata&\nodata&  14.50 & 0.20 &14.47 &14.75 \nl
 C\,{\sc iii}  &  15.50 & 0.20  &\nodata&\nodata&  15.50 & 0.20 &15.25 &15.24 \nl
 C\,{\sc iv}   &  14.41 & 0.05  & 14.40 &  0.01 &  14.00\tablenotemark{c} & 0.06 &13.99 &13.99\nl
 N\,{\sc i}    &$<$14.00&\nodata&\nodata&\nodata&$<$14.00&\nodata      &10.55 &10.97 \nl
 N\,{\sc ii}   &$<$13.70&\nodata&\nodata&\nodata&$<$13.70&\nodata      &13.00 &13.31 \nl
 N\,{\sc iii}  &  14.00 & 0.20  &\nodata&\nodata&  14.00 & 0.20 &13.79 &13.77 \nl
 N\,{\sc v}    &$<$14.00&\nodata&\nodata&\nodata&$<$14.00&\nodata      &11.30 &10.87 \nl
 O\,{\sc i}   &$<$14.10 &\nodata&\nodata&\nodata&$<$14.10&\nodata      &12.52 &12.84 \nl
 O\,{\sc vi}   &  14.95 & 0.20  &\nodata&\nodata&  14.95 & 0.20 &11.54 &10.76 \nl
 Mg\,{\sc i}   &$<$11.90&\nodata&\nodata&\nodata&$<$11.90&\nodata     &11.61 &11.68 \nl
 Mg\,{\sc ii}  &  13.28 & 0.09  &\nodata&\nodata&  13.28 & 0.09 &13.23 &13.29 \nl
 Al\,{\sc i}   &$<$12.60&\nodata&\nodata&\nodata&$<$12.60&\nodata      &10.69 &11.00 \nl
 Al\,{\sc ii}  &  12.44 & 0.03  & 12.44 &  0.01 &  12.44 & 0.03 &12.98 &13.26 \nl
 Al\,{\sc iii} &  12.45 & 0.05  & 12.39 &  0.03 &  12.45 & 0.05 &12.60 &12.86 \nl
 Si\,{\sc i}   &$<$13.20&\nodata&\nodata&\nodata&$<$13.20&\nodata      &10.29 &10.48 \nl
 Si\,{\sc ii}  &  13.52 & 0.05  & 13.52 &  0.01 &  13.52 & 0.05 &13.30 &13.47 \nl
 Si\,{\sc iii} &  14.10 & 0.20  &\nodata&\nodata&  14.10 & 0.20 &14.06 &14.12 \nl
 Si\,{\sc iv}  &\nodata &\nodata&$<$14.28&\nodata&$<$14.28&\nodata     &13.43 &13.25 \nl
 Fe\,{\sc ii}  &  12.47 & 0.08  & 11.78 &  0.12 &  12.47 & 0.08 &11.45 &11.08 \nl
\enddata					    
\tablenotetext{a}{MODEL 1: Haardt \& Madau, $\log\Gamma=-2.95$, 
$\log n_{\rm H}=-1.8$, $Z=0.63\,Z_{\rm DLA}$, $S=1.6$ kpc. }
\tablenotetext{b}{MODEL 2: Power Law $\alpha=-2.0$, $\log\Gamma=-3.02$, 
$\log n_{\rm H}=-1.91$, $Z=0.79\,Z_{\rm DLA}$, $S=1.8$ kpc. }
\tablenotetext{c}{Sum of fit components 2 and 3.} 
\end{deluxetable}				    

\clearpage

\begin{deluxetable}{lrlclclc}                             
\scriptsize        
\tablecaption{Ionic Column Densities (cm$^{-2}$) in the C\,{\sc iv} Systems 
at $z_{\rm abs}=1.66280$ toward HE 1104--1805 A 
and Photoionization Models. \label{tbl-7}}       
\tablewidth{0pt}                                                              
\tablehead{                                                                   
\colhead{Species}
&\colhead{log\,$N_{\rm fit}(X_i)$}&\colhead{$\sigma_{{\rm log}N}$}   
&\colhead{log\,$N_{\rm app}(X_i)$}&\colhead{$\sigma_{{\rm log}N}$}   
&\colhead{log\,$N_{\rm ad}(X_i)$}&\colhead{$\sigma_{{\rm log}N}$}   
&\colhead{MODEL \tablenotemark{a}}
}                                                                             
\startdata 
 H\,{\sc i}    &  16.59 & 0.07   &\nodata&\nodata &  16.05\tablenotemark{b} & 0.07 &16.05 \nl
 H\,{\sc ii}   &\nodata &\nodata &\nodata&\nodata &\nodata &      &18.92 \nl
 C\,{\sc iii}  &  13.15 & 0.20   &\nodata&\nodata &  13.15 & 0.20 &14.42 \nl
 C\,{\sc iv}   &  13.87 & 0.10   & 13.85 &  0.01  &  13.87 & 0.10 &13.84 \nl
 N\,{\sc iii}  &$<$13.00&\nodata &\nodata&\nodata &$<$13.00&      &12.97 \nl
 Mg\,{\sc ii}  &$<$11.80&\nodata &\nodata&\nodata &$<$11.80&      &11.76 \nl
 Si\,{\sc iv}  &\nodata &\nodata &$<$12.58&\nodata&$<$12.58&      &12.88 \nl
\enddata					    
\tablenotetext{a}{MODEL: Haardt \& Madau, $\log\Gamma=-2.45$, 
$\log n_{\rm H}=-2.30$, $Z=Z_{\rm DLA}$, $S=0.5$ kpc. }
\tablenotetext{b}{$N$(H\,{\sc i}) distributed according to the 
$N$(Mg\,{\sc ii}) ratios.} 
\end{deluxetable}				    

\clearpage

\begin{deluxetable}{lrlclclc}                                                  
\scriptsize                                            
\tablecaption{Ionic Column Densities (cm$^{-2}$) in the C\,{\sc iv} Systems 
at $z_{\rm abs}=1.66465$ toward HE 1104--1805 A 
and Photoionization Models. \label{tbl-8}}       
\tablewidth{0pt}                                                              
\tablehead{                                                                   
\colhead{Species}
&\colhead{log\,$N_{\rm fit}(X_i)$}&\colhead{$\sigma_{{\rm log}N}$}   
&\colhead{log\,$N_{\rm app}(X_i)$}&\colhead{$\sigma_{{\rm log}N}$}   
&\colhead{log\,$N_{\rm ad}(X_i)$}&\colhead{$\sigma_{{\rm log}N}$}   
&\colhead{MODEL \tablenotemark{a}}
}                                                                             
\startdata 
 H\,{\sc i}    &  16.59 & 0.07 &\nodata &\nodata&  16.44\tablenotemark{b} & 0.07 &16.44 \nl
 H\,{\sc ii}   &\nodata &\nodata&\nodata&\nodata&\nodata &      &19.22 \nl
 C\,{\sc iii}  &  15.42 & 0.20 &\nodata &\nodata&  15.42 & 0.20 &14.74 \nl
 C\,{\sc iv}   &  14.04 & 0.03 & 14.03  &  0.01 &  14.04 & 0.03 &14.06 \nl
 N\,{\sc iii}  &$<$13.20&\nodata&\nodata&\nodata&$<$13.20&      &13.29 \nl
 Mg\,{\sc ii}  &  12.19 & 0.03 &\nodata &\nodata&  12.19 & 0.03 &12.20 \nl
 Si\,{\sc iii} &  13.70 & 0.20 &\nodata &\nodata&  13.70 & 0.20 &13.42 \nl
 Si\,{\sc iv}  &\nodata &\nodata& 13.11 &  0.07 &$<$13.11& 0.07 &13.19 \nl
\enddata					    
\tablenotetext{a}{MODEL: Haardt \& Madau, $\log\Gamma=-2.54$, 
$\log n_{\rm H}=-2.21$, $Z=Z_{\rm DLA}$, $S=0.9$ kpc. }
\tablenotetext{b}{$N$(H\,{\sc i}) distributed according to the 
$N$(Mg\,{\sc ii}) ratios.} 
\end{deluxetable}				    

\clearpage

\begin{deluxetable}{lrlclclc}                                                  
\scriptsize                                                                 
\tablecaption{Ionic Column Densities (cm$^{-2}$) in the C\,{\sc iv} Systems 
at $z_{\rm abs}=1.66493$ toward HE 1104--1805 B 
and Photoionization Models. \label{tbl-9}}       
\tablewidth{0pt}                                                              
\tablehead{                                                                   
\colhead{Species}
&\colhead{log\,$N_{\rm fit}(X_i)$}&\colhead{$\sigma_{{\rm log}N}$}   
&\colhead{log\,$N_{\rm app}(X_i)$}&\colhead{$\sigma_{{\rm log}N}$}   
&\colhead{log\,$N_{\rm ad}(X_i)$}&\colhead{$\sigma_{{\rm log}N}$}   
&\colhead{MODEL \tablenotemark{a}}
}                                                                             
\startdata 
 H\,{\sc i}    &  15.90 &\nodata&\nodata&\nodata& 15.90  &\nodata&15.90 \nl
 H\,{\sc ii}   &\nodata &\nodata&\nodata&\nodata&\nodata &\nodata&18.90 \nl
 C\,{\sc iii}  &$<$12.90&\nodata&\nodata&\nodata&$<$12.90&\nodata&14.17 \nl
 C\,{\sc iv}   &  13.67 & 0.03  & 13.61 &  0.02 &  13.67 & 0.03 &13.67 \nl
 N\,{\sc iii}  &$<$13.20&\nodata&\nodata&\nodata&$<$13.20&\nodata&12.71 \nl
 Mg\,{\sc ii}  &$<$12.40&\nodata&\nodata&\nodata&$<$12.40&\nodata&11.33 \nl
 Si\,{\sc iv}  &\nodata &\nodata&$<$12.86&\nodata&$<$12.86&\nodata&12.58 \nl
\enddata					    
\tablenotetext{a}{MODEL: Haardt \& Madau, $\log\Gamma=-2.35$, 
$\log n_{\rm H}=-2.40$, $Z=0.63\,Z_{\rm DLA}$, $S=0.6$ kpc. }
\end{deluxetable}

\clearpage

\figcaption[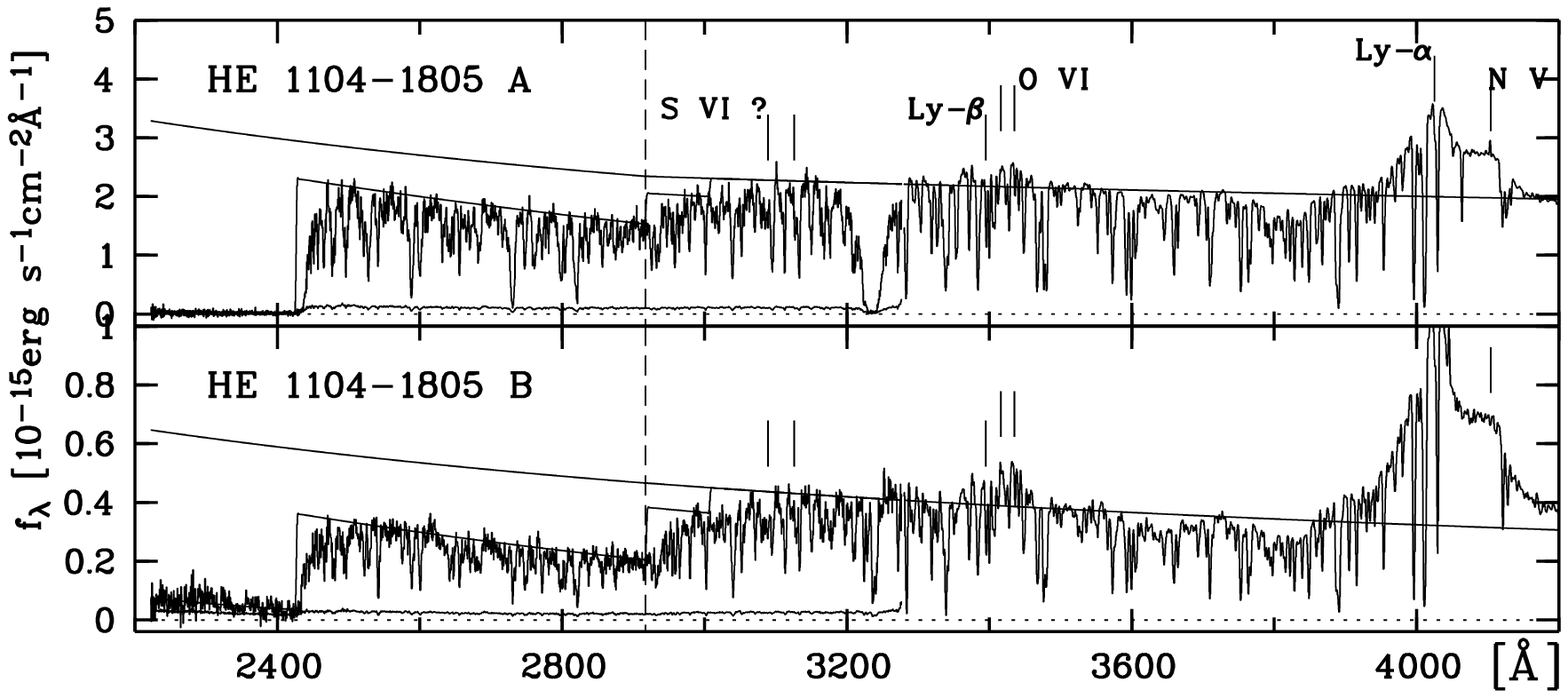]{
{\it HST} FOS spectra of HE 1104--1805 A and B and 
their 1$\sigma$ errors. Fluxes have been
corrected for galactic extinction (Seaton \cite{Seaton}) with 
E(${\rm B}-{\rm V}$)=0.09 (Reimers et al. \cite{Reimers}). 
Also shown are the {\it AAT} spectra which have been scaled to the 
{\it HST} flux levels, smoothed and rebinned to the FOS 
resolution, and plotted for $\lambda\geq$ 3277 \AA{}. 
Metal absorption lines lying longward of
Ly$\alpha$ emission were covered by the {\it NTT} and Keck spectra.
The power law and the QSO continuum derived as described in 
the text have been overplotted. Expected positions of emission lines
by Ly$\alpha$, Ly$\beta$, N\,{\sc v}, O\,{\sc vi}, and S\,{\sc vi} are
indicated.  
\label{fig1}
}

\figcaption[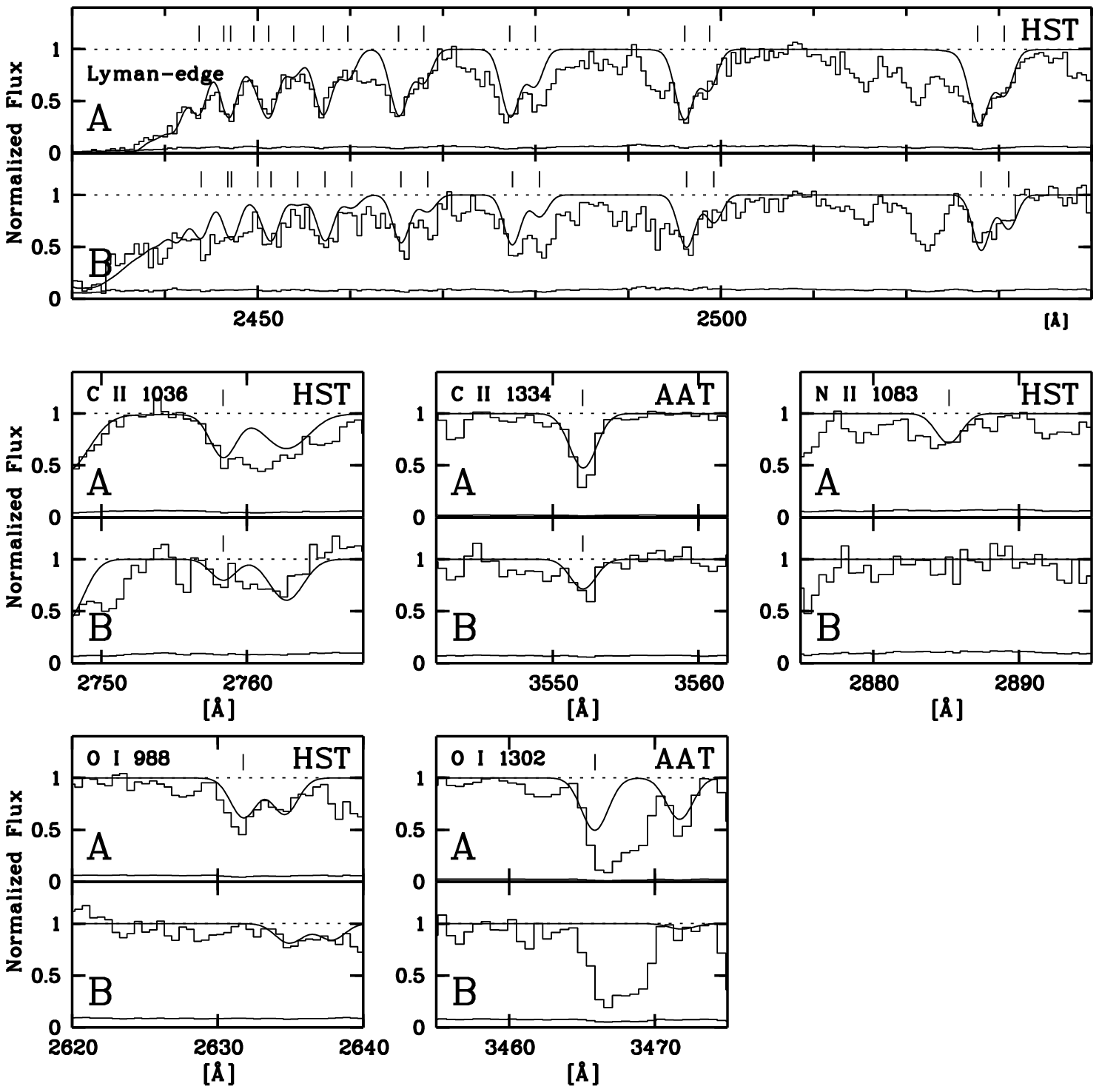]{
Sections of the normalized {\it HST} FOS (FWHM $\sim 2$ \AA) and {\it
AAT} (FWHM $\sim 1.2$ \AA) spectra of HE 1104--1805 A (upper panels) 
and B, showing synthetic Voigt profiles of the high-order H\,{\sc i}
Lyman-series lines (top) and most relevant metal lines belonging
to the damped Ly$\alpha$ system (A) and Lyman-limit System (B) at
$z=1.66$. Tickmarks (from upper panel to right-lower panel) 
indicate the positions of H\,{\sc i} $\lambda\lambda918$ to $949$, 
C\,{\sc ii} $\lambda\lambda1036,1334$, N\,{\sc ii} $\lambda1083$ and
O\,{\sc i} $\lambda\lambda988,1302$.
\label{fig8}
}

\figcaption[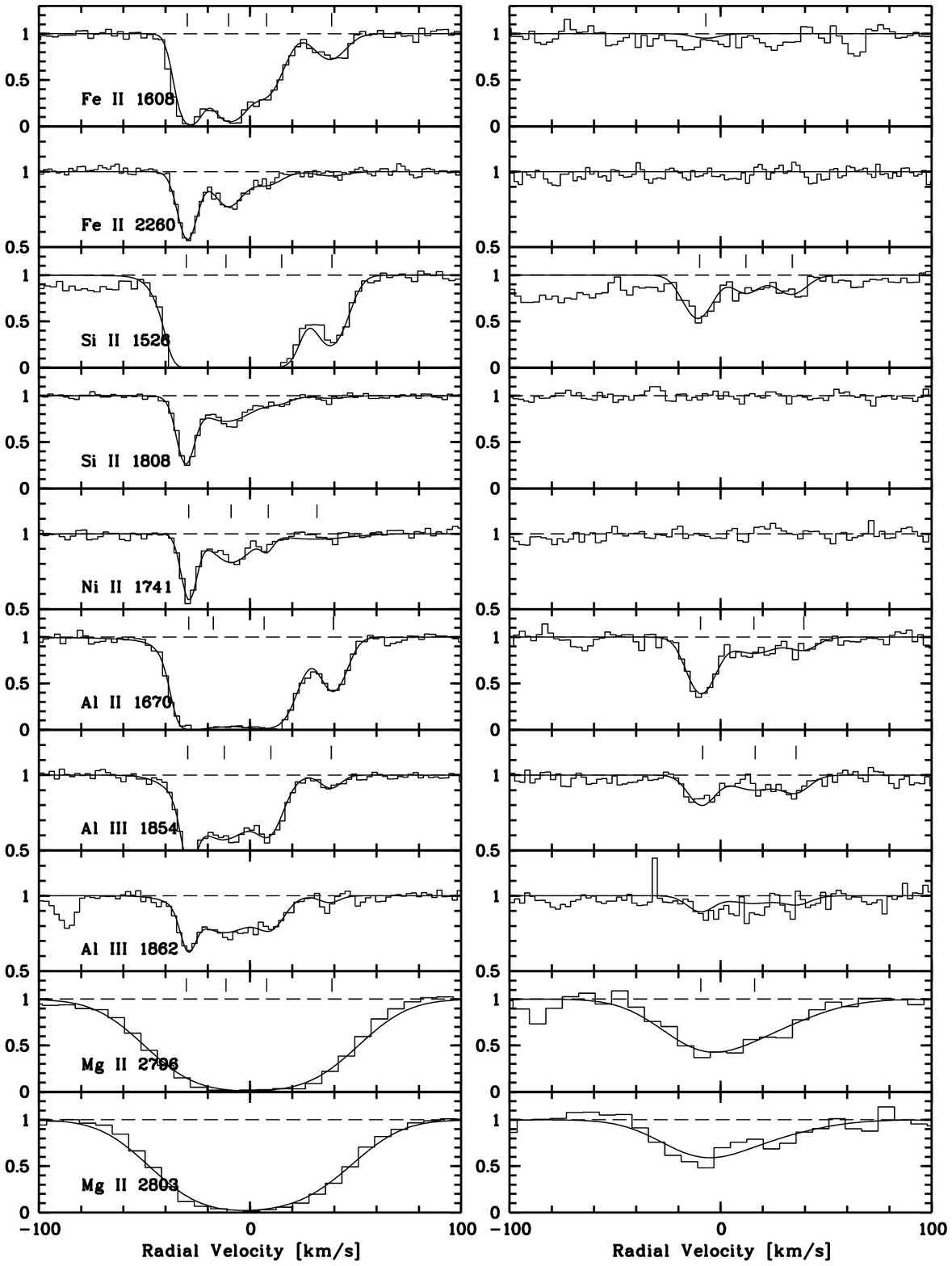]{
Low ions: Velocity profiles at 6.6 km~s$^{-1}$ resolution of
lines associated with the damped Ly$\alpha$ system 
toward HE 1104--1805 A (left-hand panel), and the Lyman-limit System
toward HE 1104--1805 B. The Mg\,{\sc ii} line profiles are at 40
km~s$^{-1}$ resolution. 
The zero velocity scale in both panels corresponds to
$z=1.66164$. The smoothed lines represent the best-fit Voigt profiles
and ticks mark the line positions. 
         \label{fig9}}

\figcaption[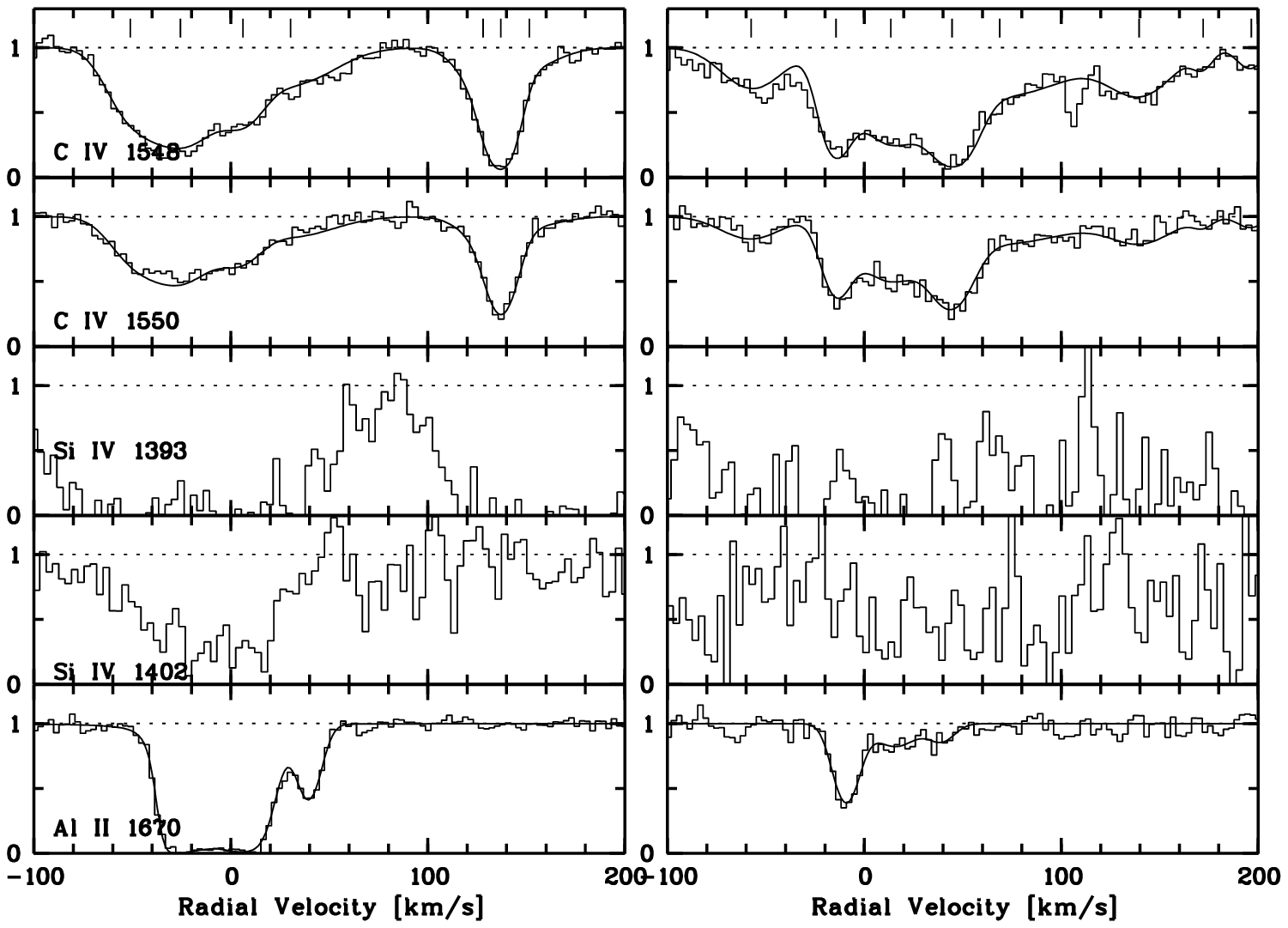]{
High ions: 
Velocity profiles at 6.6 km~s$^{-1}$ resolution of
lines associated with the damped Ly$\alpha$ system 
toward HE 1104--1805 A (left-hand panel), and the Lyman-limit System toward HE
1104--1805 B. The Al\,{\sc ii} $\lambda 1670$ velocity profiles are  
also shown for comparison purposes (see text). The zero velocity scale
in both panels corresponds to 
$z=1.66164$. The smoothed lines represent the best-fit Voigt profiles
and ticks mark the line positions. The absorption feature at $v\sim
+110$ km~s$^{-1}$ in the C\,{\sc iv} $\lambda$1548 B plot is identified
with Si\,{\sc ii} $\lambda$1808 at $z=1.28$.
         \label{fig11}}

\figcaption[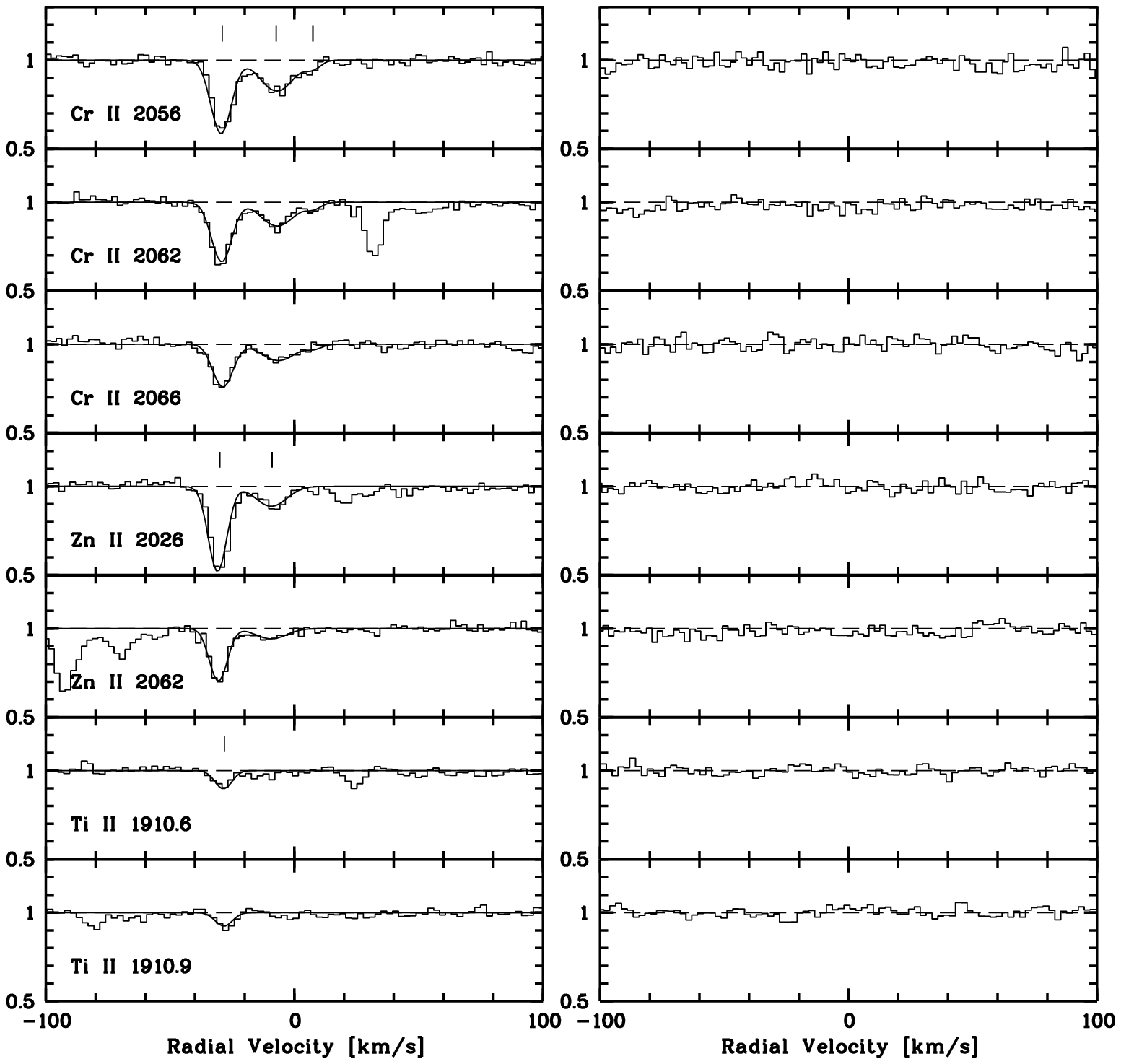]{
{\it Left:} Velocity profiles at 6.6 km~s$^{-1}$ resolution 
of Cr\,{\sc ii}, Zn\,{\sc ii} and Ti\,{\sc ii} 
lines associated with the damped Ly$\alpha$ system 
toward of HE 1104--1805 A. {\it Right:} Expected positions of the same
ions in HE 1104--1805 B. The zero velocity scale in both panels 
corresponds to $z=1.66164$. The smoothed lines represent the 
best-fit Voigt profiles and ticks mark the line positions.
\label{fig10}
}

\figcaption[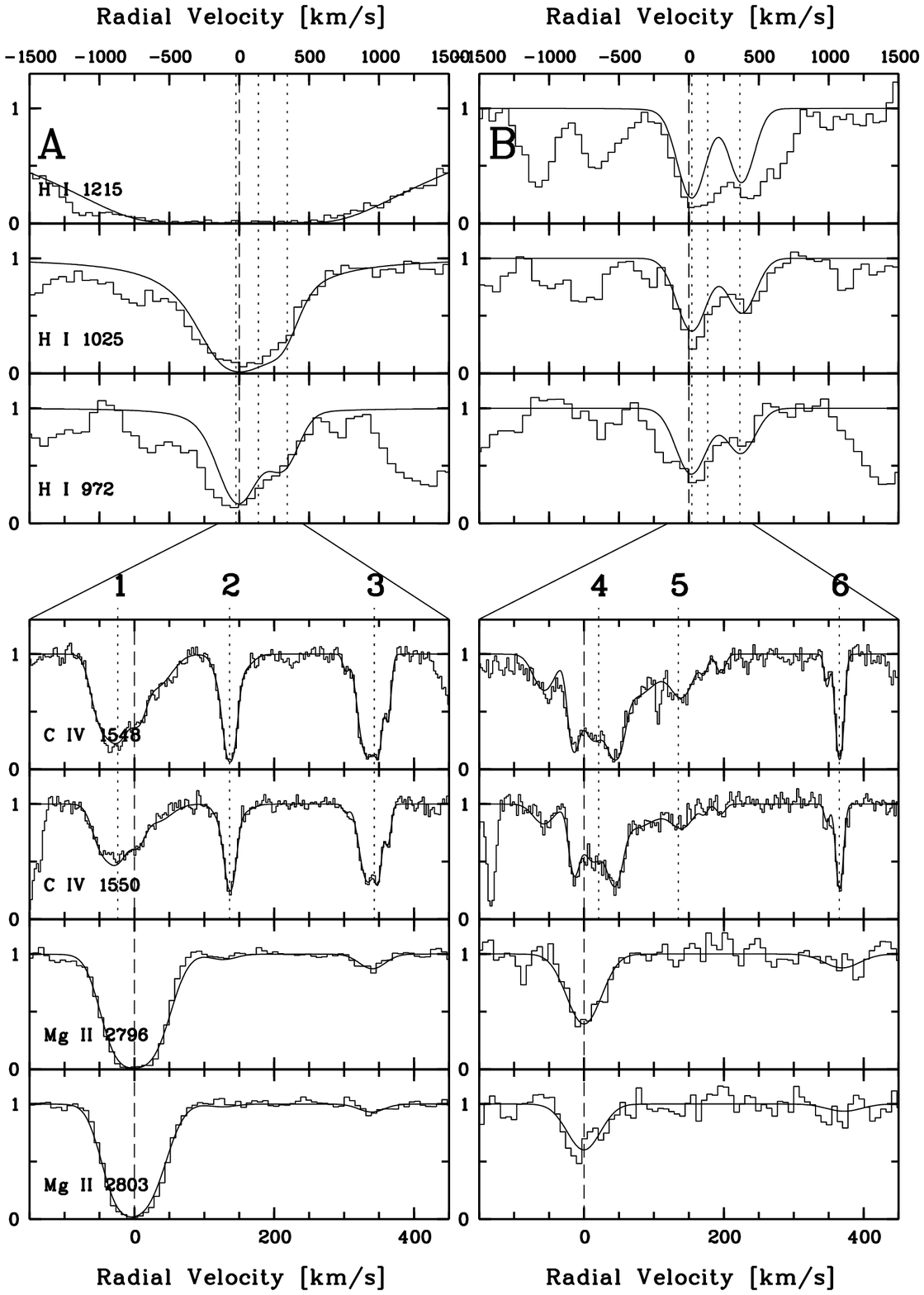]{
{\it Bottom:} Velocity profiles of the
C\,{\sc iv} $\lambda\lambda$1548,1550 (at 6.6 km~s$^{-1}$ resolution) 
and Mg\,{\sc ii} $\lambda\lambda$2796,2803 (at 40 km~s$^{-1}$ 
resolution) doublets in A (left) and B. In both panels $v=0$ km~s$^{-1}$ 
corresponds to $z=1.66164$. 
In the Mg\,{\sc ii} $\lambda$2796 plot, the feature at $-100$
km~s$^{-1}$ is identified with Mn\,{\sc i} $\lambda$2795. {\it Top:}
{\it HST} velocity profiles of the corresponding 
Ly$\alpha$, Ly$\beta$ and Ly$\gamma$ H\,{\sc i} lines 
(note the different velocity scales).
         \label{fig2}}

\figcaption[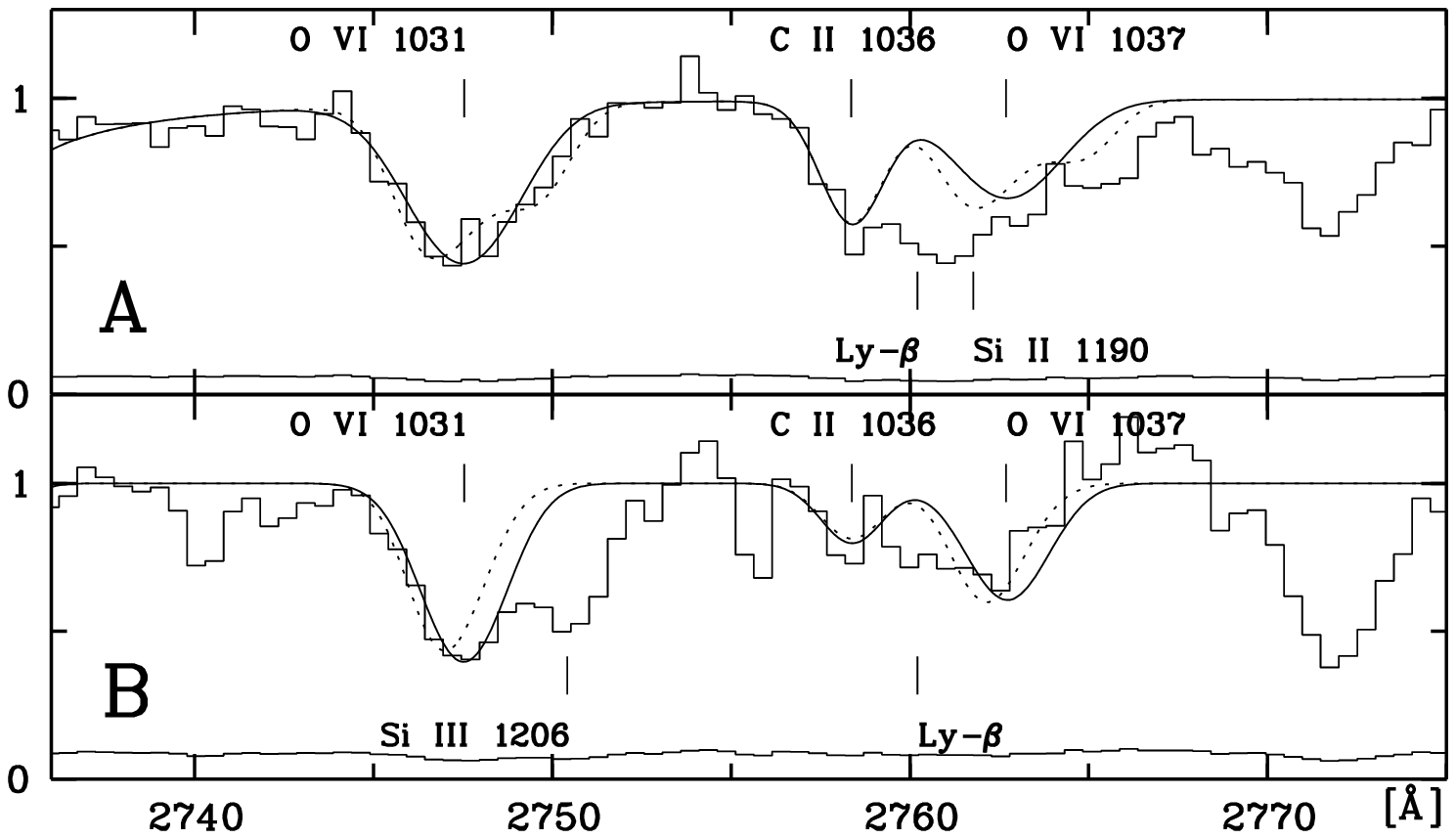]{
Section of the {\it HST} spectra A and B showing the 
O\,{\sc vi} $\lambda\lambda$1031,1037 doublet at $z$ = 1.66253. 
Solid lines: Voigt profiles---convolved with a ${\rm FWHM}=2$ \AA{}
Gaussian---with  
$b=180$ km~s$^{-1}$ (A) and $110$ km~s$^{-1}$ (B), and column density
$\log N=14.95$. Dotted lines: Voigt profiles of 3 (A) and 2 (B) 
O\,{\sc vi} lines with common $b=50$ km~s$^{-1}$ and {\it total} column
density $\log N=15.1$ (A) and $15.0$ (B). The lines are placed at
redshifts labeled 1, 2 and 3  (A), and 4 and 5 (B) in Fig.~\ref{fig2}. 
\label{fig3}
}

\figcaption[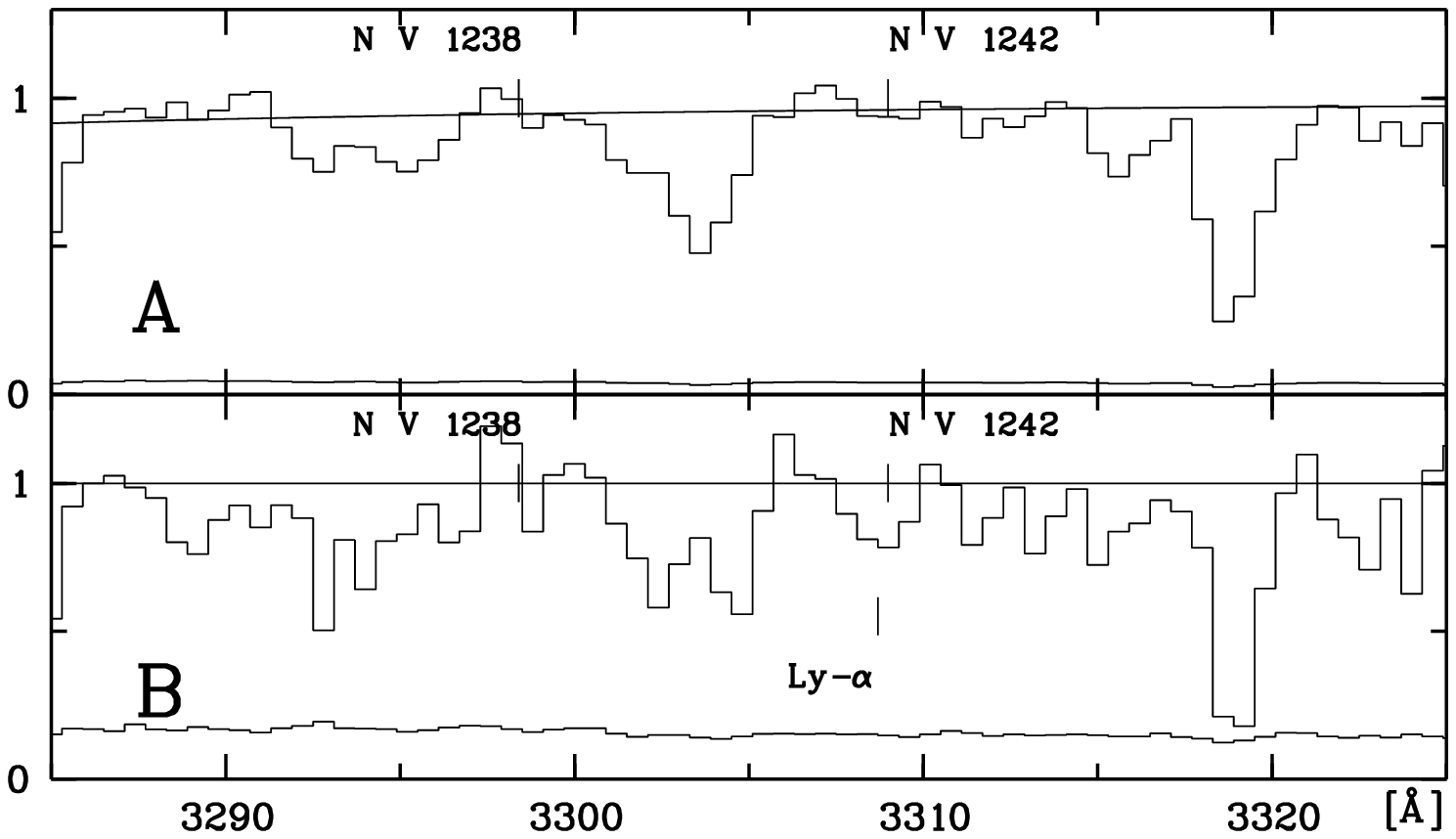]{
Section of the {\it AAT} spectra A and B. Tickmarks 
indicate the positions at which the N\,{\sc v} $\lambda\lambda1238,1242$
doublet lines  would
appear if N\,{\sc v} were present at $z=1.66253$. 
Part of the red wing of the damped Ly$\alpha$ line
appears in the overplotted synthetic spectrum A. In B, the 
absorption at $\lambda=3308$ \AA{} is identified with a Ly$\alpha$
line, supported by the detection of  Ly$\beta$ at the
same redshift in the {\it HST} spectrum.
         \label{fig4}}

\figcaption[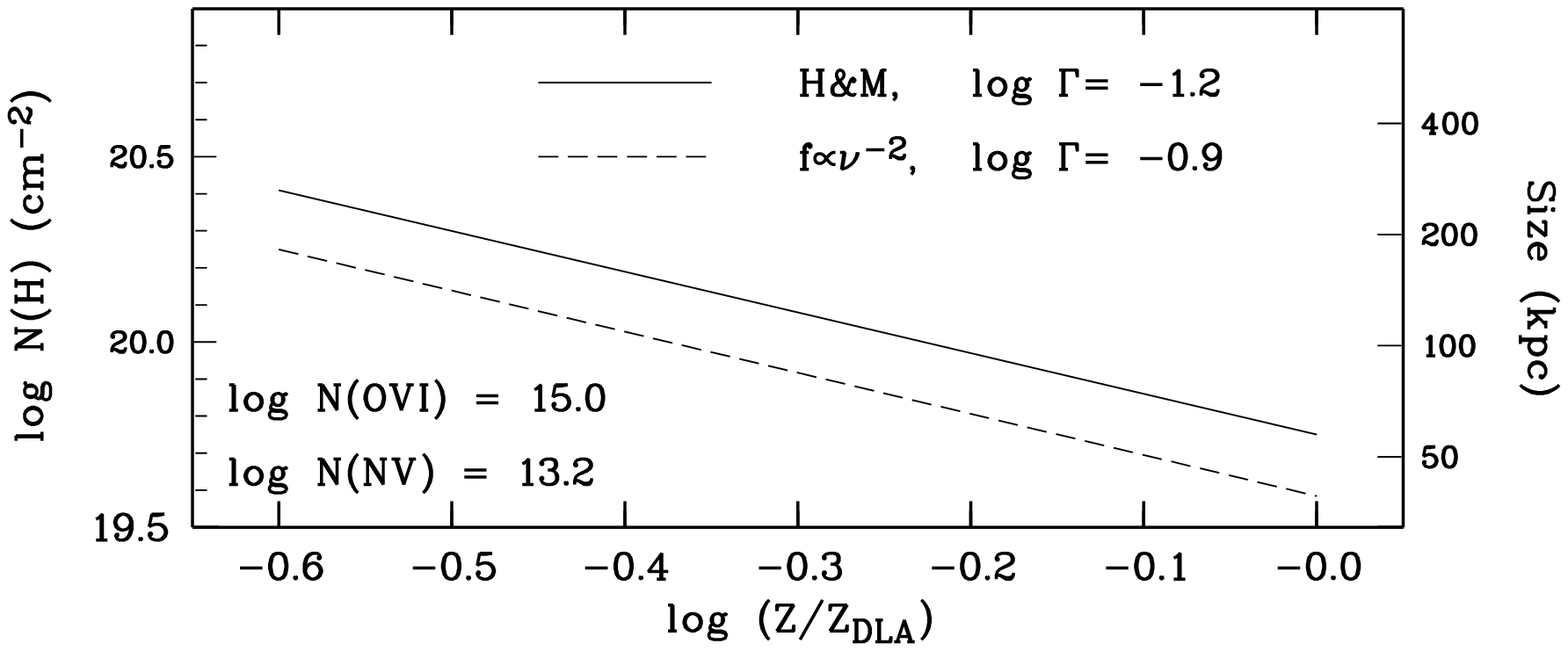]{
Hydrogen column density vs. metallicity in the O\,{\sc vi}-phase 
for different CLOUDY models assuming the Haardt \& Madau continuum 
(solid line) and a power law as ionizing background. 
The relative abundances are the same as in the DLA gas.
These models yield throughout 
$\log N$(O\,{\sc vi})$=15.0$ and $\log N$(N\,{\sc v})$=13.2$. 
         \label{fig5}}

\figcaption[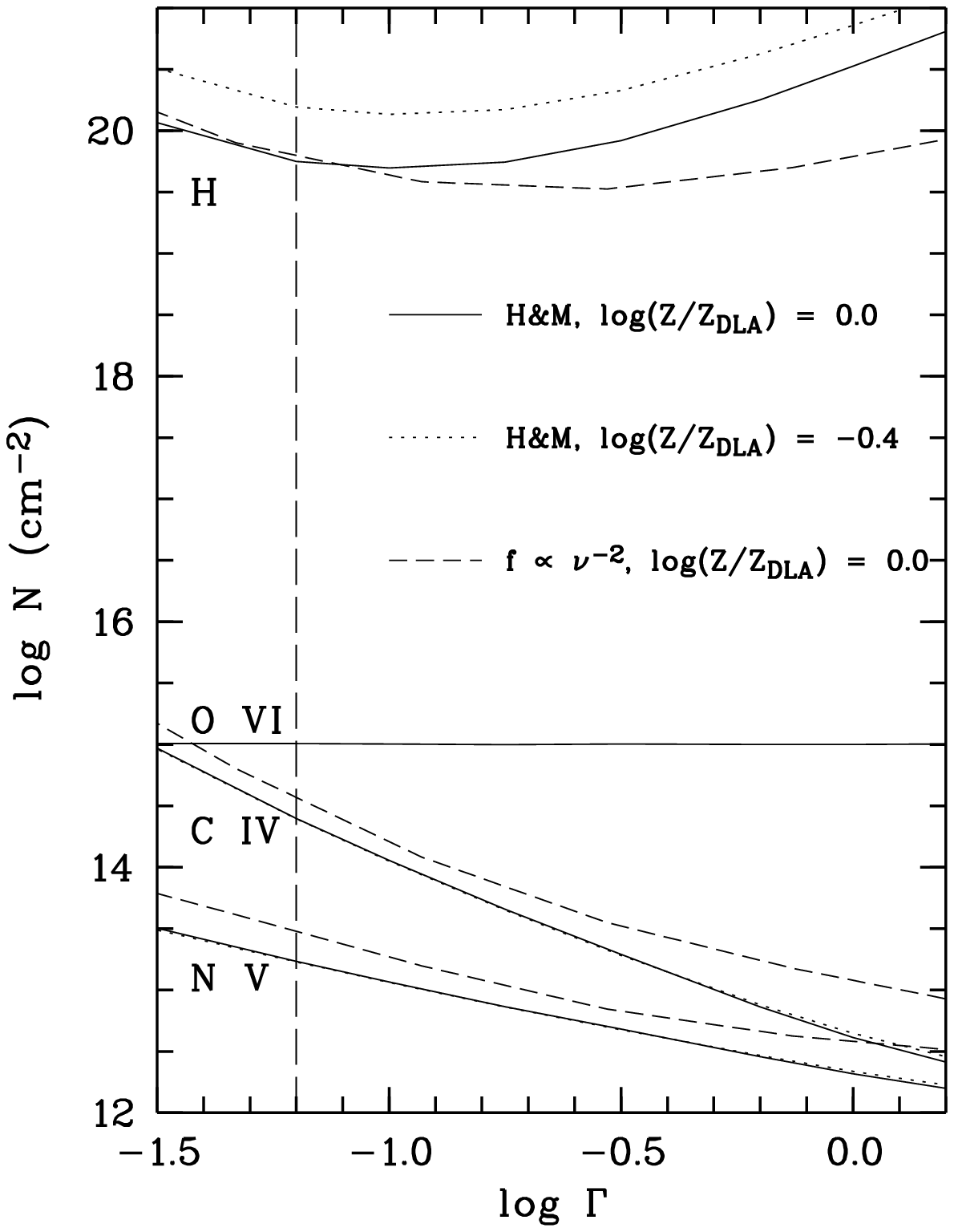]{
Column densities vs. ionization parameter $\Gamma$ in the O\,{\sc
vi}-phase for different CLOUDY models. {\it Solid line:} Haardt \&
Madau radiation field as ionizing background, and $Z=Z_{\rm
DLA}$. {\it Dotted line:} Same radiation field, with $Z=0.4~Z_{\rm
DLA}$. {\it Dashed line:} Power law  and  $Z=Z_{\rm DLA}$. 
The relative abundances are the same as in the DLA gas. 
These models yield throughout $\log N$(O\,{\sc vi})$=15.0$. 
         \label{fig12}}

\clearpage
\plotone{f1.ps}
\clearpage
\plotone{f2.ps}
\clearpage
\plotone{f3.ps}
\clearpage
\plotone{f4.ps}
\clearpage
\plotone{f5.ps}
\clearpage
\plotone{f6.ps}
\clearpage
\plotone{f7.ps}
\clearpage
\plotone{f8.ps}
\clearpage
\plotone{f9.ps}
\clearpage
\plotone{f10.ps}


\clearpage

\begin{thebibliography}{}

    \bibitem[1993]{Bahcall}
      Bahcall, J. N., et al. 1993, \apjs, 87, 1 
    \bibitem[1986]{Bergeron1}
      Bergeron, J., \& Stasinska, G. 1986, \aap, 169, 1 
    \bibitem[1994]{Bergeron}
      Bergeron, J., et al. 1994, \apj, 436, 33 
    \bibitem[1996]{Burles}
      Burles, S., \& Tytler, D. 1996, \apj, 460, 584 
    \bibitem[1991]{Cardelli}
      Cardelli, J. A., Savage, B. D., \& Ebbets, D. C. 1991, \apj, 383,
      L23 
    \bibitem[1995]{Cardelli2}
      Cardelli, J. A., Sembach, K. R. \& Savage, B. D. 1995, \apj, 440,
      241 
    \bibitem[1989]{Caulet}
      Caulet, A. 1989, \apj, 340, 90 
    \bibitem[1998]{Courbin}
      Courbin, F., Lidman, C. \& Magain, P. 1998, \aap, 330, 57 
    \bibitem[1993]{Ferland}
      Ferland, G. J. 1993, University of Kentucky, Physics Department
      Internal Report
    \bibitem[1995]{Fontana}
      Fontana, A. \& Ballester, P. 1995, {\it ESO Messenger} 80, 37
    \bibitem[1996]{Haardt} 
      Haardt, F., \& Madau, P. 1996, \apj, 461, 20
    \bibitem[1996a]{Haehnelt1} 
      Haehnelt, M. G., Rauch, M., \& Steinmetz, M. 1996a, \mnras, 283, 1055
    \bibitem[1996b]{Haehnelt} 
      Haehnelt, M. G., Steinmetz, M. \& Rauch, M. 1996b, \apj, 465, L95
    \bibitem[1998]{Haehnelt2} 
      Haehnelt, M. G., Steinmetz, M. \& Rauch, M. 1998, \apj, 495, 647
    \bibitem[1997]{Kirkman} 
      Kirkman, D., \& Tytler, D. 1997, \apj, 489, L123
    \bibitem[1993]{Lu}
      Lu, L., \& Savage, B. D. 1993, \apj, 403, 127
    \bibitem[1996]{Lu1}
      Lu, L., Sargent, W. L. W., Barlow, T. A., Churchill, C. W., \&
      Vogt, S. S. 1996, \apjs, 107, 475
    \bibitem[1995]{Mar} Mar, D. P., \& Bailey, G. 1995, PASA, 12, 239
    \bibitem[1991]{Morton}
      Morton, D. C. 1991, \apjs, 77, 119
    \bibitem[1990]{Pettini3}
      Pettini, M., Boksenberg, A., \& Hunstead, R.W. 1990,
      \apj, 348, 48
    \bibitem[1995]{Pettini1}
      Pettini, M., Lipman, K., \& Hunstead, R.W. 1995,
      \apj, 451, 100
    \bibitem[1997]{Pettini2}
      Pettini, M., Smith, L. J., Kink, D. L., \& Hunstead, R.W. 1997,
      \apj, 486, 665
    \bibitem[1986]{Press} 
      Press, W. H., Flannery, B. P., Teukolsky, S. A., \& Vetterling,
      W. T. 1986, Numerical Recipes, Cambridge University Press.
    \bibitem[1996]{Prochaska1} 
      Prochaska, J. X., \& Wolfe, A. M. 1996, \apj, 470, 403
    \bibitem[1997]{Prochaska2} 
      Prochaska, J. X., \& Wolfe, A. M. 1997, \apj, 474, 140
    \bibitem[1997a]{Rauch} 
      Rauch, M., Haehnelt, M. G., \& Steinmetz, M. 1997a, \apj, 481, 601
    \bibitem[1997b]{Rauch1} 
      Rauch, M. 1997b, in {\it Structure and Evolution of the IGM from
      QSO Absorption Lines}, Proc. 13th IAP Colloquium,
      ed. P. Petitjean, S. Charlot (Paris: Editions Fronti\`eres), 109  
    \bibitem[1995]{Reimers} 
      Reimers, D., Rodriguez-Pascual, P., Hagen, H.-J., \& Wisotzi,
      L. 1995, \aap, 293, L21
    \bibitem[1998]{Remy} 
      Remy, M., Claeskens, J.-F., Surdej, J., et al. 1998,
      {\it New Astronomy}, in press.
    \bibitem[1991]{Savage1} 
      Savage, B. D. \&{} Sembach, K. R. 1991, \apj, 379, 245
    \bibitem[1996]{Savage} 
      Savage, B. D. \&{} Sembach, K. R. 1996, \apj, 470, 893
    \bibitem[1993]{Schneider} 
      Schneider, D. P., et al. 1993, \apjs, 87, 45
    \bibitem[1979]{Seaton} 
      Seaton, M. J. 1979, \mnras, 187, 73
    \bibitem[1996]{Sembach} 
      Sembach, K. R. \& Savage, B. D. 1996, \apj, 457, 211
    \bibitem[1992]{Smette1} 
      Smette, A., Surdej, J., Shaver, P. A., Foltz, C. B., Chaffee,
      F. H., Weymann, R. J., Williams, R. E. \& Magain, P. 1992, \apj, 389, 39
    \bibitem[1995]{Smette} 
      Smette, A., Robertson, J. G., Shaver, P. A., Reimers, D., Wisotzki,
      L. \& K\"ohler T. (Paper I) 1995, \aaps, 113, 199
    \bibitem[1993]{Sutherland} 
      Sutherland, R. S. \& Dopita, M. A. 1993, \apjs, 88, 253
    \bibitem[1994]{Verner}
      Verner, D. A., Barthel, P. D. \& Tytler, D. 1994, \aaps, 108, 287 
    \bibitem[1998]{Vladilo}
      Vladilo, G., 1998, \apj, 493, 583
    \bibitem[1994]{Vogt}
      Vogt, S. S., Allen, S. L., Bigelow, B. C., Bresee, L., Brown, B.,
      et al. 1994, S.P.I.E. 2198:362
    \bibitem[1993]{Wisotzki1}
      Wisotzki, L., K\"ohler, T., Kayser, R., \& Reimers, D. 1993, \aap,
      278, L15
    \bibitem[1995]{Wisotzki2}
      Wisotzki, L., K\"ohler, T., Ikonomou, M., \& Reimers, D. 1995,
      \aap, 297, L59
    \bibitem[1995a]{Wolfe} 
      Wolfe, A. M. 1995a, in {\it QSO Absorption Lines}, Proc. ESO
      Workshop, ed. G. Meylan (Heidelberg: Springer), 13 
    \bibitem[1995b]{Wolfe1}
      Wolfe, A. M., Lanzetta, K. M., Foltz, C. B., \& Chaffee F. H. 1995b,
      \apj, 454, 698
    \bibitem[1998]{Wolfe2} 
      Wolfe, A. M., \& Prochaska, J. X. 1998, \apj, 494, L15
    \bibitem[1997]{Zuo}
      Zuo, L., Beaver, E. A., Burbidge, E. M., Cohen, R. D.,
      Junkkarinen, V. T., \& Lyons, R. W. 1997, \apj, 477, 568

\end{thebibliography}
\end{document}